\documentclass[twocolumn,showpacs,preprintnumbers,amsmath,amssymb]{revtex4}

\usepackage{graphicx}% Include figure files
\usepackage{dcolumn}% Align table columns on decimal point
\usepackage{bm}% bold math

\normalsize

\begin{document}

\title{Storing and processing optical information with ultra-slow light in Bose-Einstein condensates}

\author{Zachary Dutton$^{1,2}$ and Lene Vestergaard Hau$^2$}
\affiliation{ ${}^1$National Institute of Standards \& Technology,
Electron and
Optical Division, Gaithersburg MD 20899-8410 \\
${}^2$Department of Physics and
Division of Engineering and Applied Sciences, \\
Harvard University, Cambridge MA 02138}

\date{\today}

\begin{abstract}
We theoretically explore coherent information transfer between
ultra-slow light pulses and Bose-Einstein condensates (BECs) and
find that storing light pulses in BECs allows the coherent
condensate dynamics to process optical information. We consider
BECs of alkali atoms with a $\Lambda$ energy level configuration.
In this configuration, one laser (the coupling field) can cause a
pulse of a second pulsed laser (the probe field) to propagate with
little attenuation (electromagnetically induced transparency) at a
very slow group velocity ($\sim 10$~m/s), and be spatially
compressed to lengths smaller than the BEC.  These pulses can be
fully stopped and later revived by switching the coupling field
off and on. Here we develop a formalism, applicable in both the
weak and strong probe regimes, to analyze such experiments and
establish several new results: (1) We show that the switching can
be performed on time scales much faster than the adiabatic time
scale for EIT, even in the strong probe regime.  We also study the
behavior of the system changes when this time scale is faster than
the excited state lifetime. (2) Stopped light pulses write their
phase and amplitude information onto spatially dependent atomic
wavefunctions, resulting in coherent two-component BEC dynamics
during long storage times. We investigate examples relevant to
Rb-87 experimental parameters and see a variety of novel dynamics
occur, including interference fringes, gentle breathing
excitations, and two-component solitons, depending on the relative
scattering lengths of the atomic states used and the probe to
coupling intensity ratio.  We find the dynamics when the levels
$|F=1, M_F=-1 \rangle$ and $|F=2, M_F=+1 \rangle$ are used could
be well suited to designing controlled processing of the
information input on the probe. (3) Switching the coupling field
on after the dynamics writes the evolved BEC wavefunctions density
and phase features onto a revived probe pulse, which then
propagates out. We establish equations linking the BEC
wavefunction to the resulting output probe pulses in both the
strong and weak probe regimes. We then identify sources of
deviations from these equations due to absorption and distortion
of the pulses.  These deviations result in imperfect fidelity of
the information transfer from the atoms to the light fields and we
calculate this fidelity for Gaussian shaped features in the BEC
wavefunctions. In the weak probe case, we find the fidelity is
affected both by absorption of very small length scale features
and absorption of features occupying regions near the condensate
edge. We discuss how to optimize the fidelity using these
considerations.  In the strong probe case, we find that when the
oscillator strengths for the two transitions are equal the
fidelity is {\it not} strongly sensitive to the probe strength,
while when they are unequal the fidelity is worse for stronger
probes. Applications to distant communication between BECs,
squeezed light generation and quantum information are anticipated.
\end{abstract}


\maketitle

\section{\label{sec:intro} Introduction}

In discussions of quantum information technology
\cite{quantumInfo}, it has been pointed out that the ability to
coherently transfer information between ``flying'' and stationary
qubits will be essential. Atomic samples are good candidates for
quantum storage and processing due to their long coherence times
and large, controllable interactions, while photons are the
fastest and most robust way to transfer information.  This implies
that methods to transfer information between atoms and photons
will be important to the development of this technology.

\begin{figure}
\includegraphics{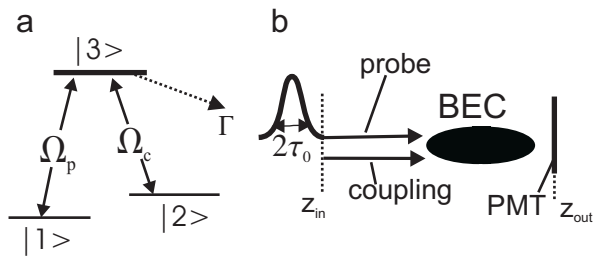}
\caption{\label{fig:diagram}\textbf{Schematic of ultra-slow light
(USL) and stopped experiments.} \textbf{(a)} The $\Lambda$ energy
level structure which we consider has two stable states, $|1
\rangle$ and $|2 \rangle$, and an excited state $|3\rangle$ which
decays at a rate $\Gamma$ (which is $(2 \pi)~10$~MHz for Na and
$(2 \pi)~6$~MHz for Rb-87). Atoms which spontaneously decay from
$|3 \rangle$ are assumed to exit the levels under consideration.
\textbf{(b)} We consider the two light fields to be {\it
co-propagating} and input in the $+z$ direction (the long axis of
the Bose-Einstein condensate (BEC)) at $z_{\mathrm{in}}$. The
output intensity at $z_{\mathrm{out}}$ can be detected
experimentally with a photo-multiplier tube (PMT).  The probe
field is pulsed with half-width $\tau_0$. The coupling beam is
c.w. but can be switched off quickly to stop the probe pulse while
it is in the BEC. Switching it back on later regenerates the probe
light pulse.}
\end{figure}

Recently the observation of ultra-slow light (USL)
\cite{Nature1,OtherUSL}, propagating at group velocities more than
seven orders of magnitude below its vacuum speed ($V_g \sim
10^{-7}c$), and the subsequent stopping and storing of light
pulses in atomic media \cite{Nature2,OtherStoppedLight} has
demonstrated a tool to possibly accomplish this
\cite{quantumProcessing}. The technique relies on the concept of
{\it electromagnetically induced transparency} (EIT)\cite{EIT} in
three level $\Lambda$-configuration atoms (see
Fig.~\ref{fig:diagram}(a)).  A coupling light field $\Omega_c$ is
used to control the propagation of a pulse of probe light
$\Omega_p$.  The probe propagates at a slow group velocity and, as
it is doing so, coherently imprints it's amplitude and phase on
the coherence between two stable internal states of the atoms,
labelled $|1 \rangle$ and $|2 \rangle$ (which are generally
particular hyperfine and Zeeman sublevels). Switching the coupling
field off stops the probe pulse and ramps it's intensity to zero,
freezing the probe's coherent (that is, intensity {\it and} phase)
information into the atomic media, where it can be stored for a
controllable time. Switching the coupling field back on at a later
time writes the information back onto a revived probe pulse, which
then propagates out of the atom cloud and can be detected, for
example, with a photo-multiplier tube (PMT) (see
Fig.~\ref{fig:diagram}(b)). In the original experiment
\cite{Nature2}, the revived output pulses were indistinguishable
in width and amplitude from non-stored USL pulses, indicating the
switching process preserved the information in the atomic medium
with a high fidelity.

In addition to the ability to store coherent information,
compressing ultra-slow probe pulses to lengths shorter than the
atomic cloud puts the atoms in spatially dependent superpositions
of $|1 \rangle$ and $|2 \rangle$, offering novel possibilities to
study two component BEC dynamics \cite{HauTalks}. In
\cite{SolitonVortex} slow light pulses propagating in
Bose-Einstein condensates (BECs) \cite{BEC} of sodium were used to
create density defects with length scales $\sim 2~\mu \mathrm{m}$,
near the BEC healing length, leading to quantum shock waves and
the nucleation of vortices.

Motivated by both observations of coherent optical information
storage in atom clouds and the interesting dynamics which slow and
stopped light pulses can induce in BECs, the present paper
theoretically examines new possibilities offered by stopping light
in BECs. We first present a novel treatment of the switching
process and establish that the switch-off and switch-on of the
coupling field can occur without dissipating the coherent
information, provided the length scale of variations in the atomic
wavefunctions are sufficiently large. We find that the switching
of the coupling field can be done {\it arbitrarily} fast in the
ultra-slow group velocity limit $V_g \ll c$. This is consistent
with previous work
\cite{Nature2,stoppedLight,fastStopped,stoppedTheory}, but goes
further as it applies to pulses in both the weak and strong probe
regimes \cite{strongStopped}.  We also present the first explicit
calculation of how the behavior of the system differs when one
switches the coupling field faster than the natural lifetime of
the excited state $\Gamma^{-1}$ (see Fig.~\ref{fig:diagram}(a))
and find the information is successfully transferred even in this
regime. Furthermore we will see the analysis presented here, being
phrased in terms of the spatial characteristics of the atomic
wavefunctions, is well suited to addressing the issue of storage
for times long compared to the time scale for atomic dynamics.

Investigation of this very issue of longer storage times
constitute the central results of the paper.  In this regime,
there are important differences between optical information
storage in BECs versus atom clouds above the condensation
temperature (thermal clouds), which have been used in experiments
to date. During the storage time, the light fields are off and the
atomic wavefunctions evolve due to kinetic energy, the external
trapping potential, and atom-atom interactions. (We label all
these {\it external} dynamics to distinguish them from couplings
between the internal levels provided by the light fields.)  To
successfully regenerate the probe pulse by then switching on the
coupling field, there must be a non-zero value of the {\it
ensemble average} (over all atoms) of the coherence between $|1
\rangle$ and $|2 \rangle$. When revivals were attempted after
longer storage times in thermal clouds (several milliseconds in
\cite{Nature2}) the external dynamics had washed out the phase
coherence between $|1 \rangle$ and $|2 \rangle$ (due to the atoms
occupying a distribution of single particle energy levels)  and no
output pulses were observed. In zero temperature BECs the
situation is completely different.  There each atom evolves in an
identical manner, described by a non-linear Schrodinger equation
(Gross-Pitaevskii equation \cite{BEC}), preserving the ensemble
average of the coherence during the external dynamics. That is,
even as the amplitude and phase of the wavefunctions representing
the BEC evolve, the relative phase of the two components continues
to be well defined at all points in space.  Thus, if we switch the
coupling field back on, we expect that the evolved BEC
wavefunctions will be written onto a revived probe field.  In this
way, the probe pulses can be {\it processed} by the BEC dynamics.

We show several examples, relevant to current Rb-87 BECs, of the
interesting two component dynamics which can occur during the
storage time. This atom possesses internal states with very small
inelastic loss rates \cite{inelastic}, and thus, long lifetimes
\cite{multipleComponentJILA}. We find that, depending upon the
relative scattering lengths of the internal states involved and
the relative intensity of the probe and coupling fields, one could
observe the formation of interference fringes, gentle breathing
motion, or the formation and motion of two-component (vector)
solitons \cite{twoCompSolitons}.  In particular we find that using
the levels $|F=1,M_F=-1 \rangle$ and $|F=2,M_F=+1 \rangle$ in
Rb-87 could allow for long, robust storage of information and
controllable processing.  In each case, we observe the amplitude
and phase due to the dynamical evolution is written onto revived
probe pulses. These pulses then propagate out of the BEC as slow
light pulses, at which point they can be detected, leaving behind
a BEC purely in its original state $|1 \rangle$.

For practical applications of this technique to storage and
processing, one must understand in detail how, and with what
fidelity, the information contained in the atomic coherence is
transferred and output on the light fields.  Thus, the last part
of the paper is devoted to finding the exact relationship between
the BEC wavefunctions before the switch-on and the observed output
probe pulses.   We find equations linking the two in the ideal
limit (without absorption or distortion).  Then, using our earlier
treatment of the switching process, we identify several sources of
imperfections and calculate the fidelity of the information
transfer for Gaussian shaped wavefunctions of various amplitudes
and lengths. In the weak probe limit, we find a simple
relationship between the wavefunction in $|2 \rangle$ and the
output probe field.  We find that optimizing the fidelity involves
balancing considerations related to, on one hand, absorption of
small length scale features in the BEC wavefunctions and, on the
other hand, imperfect writing of wavefunctions which are too near
the condensate edge. For stronger probes, we find a more
complicated relationship, though we see one can still reconstruct
the amplitude and phase of the wavefunction in $|2\rangle$ using
only the output probe.  In this regime, we find that the fidelity
of the writing and output is nearly independent of the probe
strength when the oscillator strengths for the two transitions
involved ($|1 \rangle \leftrightarrow |3 \rangle$ and $|2 \rangle
\leftrightarrow |3 \rangle$) are equal. However, unequal
oscillator strengths lead to additional distortions and phase
shifts, and therefore lower fidelity of information transfer, for
stronger probes.

We first, in Section~\ref{sec:formalism}, introduce a formalism
combining Maxwell-Schrodinger and Gross-Pitaevskii equations
\cite{BEC}, which self-consistently describe both the atomic
(internal and external) dynamics as well as light field
propagation.  We have written a code which implements this
formalism numerically. Section~\ref{sec:switching} presents our
novel analysis of the switching process, whereby the coupling
field is rapidly turned off or on, and extends previous treatments
of the fast switching regime, as described above.
Section~\ref{sec:processing} shows examples of the very rich
variety of two-component dynamics which occur when one stops light
pulses in BECs and waits for much times longer than the
characteristic time scale for atomic dynamics.
Section~\ref{sec:systematic} contains our quantitative analysis of
the fidelity with which the BEC wavefunctions are transferred onto
the probe field. We conclude in Section~\ref{sec:conclusion} and
anticipate how the method studied here could eventually be applied
to transfer of information between distant BECs, and the
generation of light with squeezed statistics.

\section{\label{sec:formalism} Description of ultra-slow light
in Bose-Einstein condensates}

We first introduce our formalism to describe the system and review
USL propagation within this formalism.  Assume a Bose-condensed
sample of alkali atoms, with each atom in the BEC containing three
internal (electronic) states in a $\Lambda$-configuration
(Fig.~\ref{fig:diagram}(a)).  The states $|1 \rangle$ and $|2
\rangle$ are stable and the excited level $|3 \rangle$,
radiatively decays at $\Gamma$.  In the alkalis which we consider,
these internal levels correspond to particular hyperfine and
Zeeman sublevels for the valence electron.   When the atoms are
prepared in a particular Zeeman sub-level and the proper light
polarizations and frequencies are used
\cite{Nature2,SolitonVortex}, this three level analysis is a good
description in practice.  All atoms are initially condensed in $|1
\rangle$ and the entire BEC is illuminated with a {\it coupling
field}, resonant with $|2 \rangle \leftrightarrow |3 \rangle$
transition and propagating in the $+z$ direction (see
Fig.~\ref{fig:diagram}(b)). A pulse of {\it probe field}, with
temporal half-width $\tau_0$, resonant with the $|1 \rangle
\leftrightarrow |3 \rangle$ transition, and also propagating in
the $+z$ direction, is then injected into the medium. The presence
of the coupling field completely alters the optical properties of
the atoms as seen by the probe. What would otherwise be an opaque
medium (typical optical densities of BECs are $\sim 400$) is
rendered transparent via {\it electromagnetically induced
transparency} (EIT) \cite{EIT}, and the light pulse propagates at
ultra-slow group velocities ($\sim 10$~m/s)\cite{Nature1}. As this
occurs, all but a tiny fraction of the probe energy is temporarily
put into the coupling field, leading to a compression of the probe
pulse to a length smaller than the atomic medium itself
\cite{Nature1,Nature2,HarrisHau}.

To describe the system theoretically, we represent the probe and
coupling electric fields with their Rabi-frequencies
$\Omega_{p(c)}=- \mathbf{d}_{13(23)} \cdot
\mathbf{E}_{p(c)}/\hbar$, where $\mathbf{E}_{p(c)}$ are slowly
varying envelopes of the electric fields (both of which can be
time and space-dependent) and $\mathbf{d}_{13(23)}$ are the
electric dipole moments of the transitions.  The BEC is described
with a two-component spinor wavefunction $(\psi_1,\psi_2)^T$
representing the mean field of the atomic field operator for
states $|1 \rangle$ and $|2 \rangle$.  We ignore quantum
fluctuations of these quantities, which is valid when the
temperature is substantially below the BEC transition temperature
\cite{FiniteTemp}.  The excited level $|3 \rangle$ can be
adiabatically eliminated \cite{adiabatic} when the variations of
the light fields' envelopes are slow compared to the excited state
lifetime $\Gamma^{-1}$ (which is 16~ns in sodium). The procedure
is outlined in the Appendix.   The functions $\psi_1,\psi_2$
evolve via two coupled Gross-Pitaevskii (GP) equations \cite{BEC}.
For the present paper, we will only consider dynamics in the $z$
dimension, giving the BEC some cross-sectional area $A$ in the
transverse dimensions over which all dynamical quantities are
assumed to be homogeneous. This model is sufficient to demonstrate
the essential effects here. We have considered effects due to the
transverse dimensions, but a full exploration of these issue is
beyond our present scope. The GP equations are
\cite{SolitonVortex,thesis}:

\begin{eqnarray}
\label{eq:formalism12} i \hbar \frac{\partial \psi_1 }{\partial t}
& = & \bigg[-\frac{\hbar^2}{2m}\frac{\partial^2}{\partial z^2} +
V_1(z) + U_{11} | \psi_1 |^2   +  U_{12}| \psi_2 |^2 \bigg] \psi_1  \nonumber \\
& & - i \hbar \left( \frac{|\Omega_p|^2}{2 \Gamma} \psi_1
+\frac{\Omega_p^\ast \Omega_c}{2 \Gamma} \psi_2 \right), \nonumber \\
i \hbar \frac{\partial \psi_2}{\partial t} & = &
\bigg[-\frac{\hbar^2}{2m}\frac{\partial^2}{\partial z^2} + V_2(z)
+ U_{22}
|\psi_2|^2 +  U_{12}| \psi_1 |^2 \bigg] \psi_2 \nonumber \\
& &  - i \hbar \left( \frac{|\Omega_c|^2}{2 \Gamma} \psi_2 +
\frac{\Omega_p \Omega_c^\ast}{2 \Gamma} \psi_1 \right),
\end{eqnarray}

\noindent where $m$ is the mass of the atoms and we will consider
a harmonic external trapping potential $V_1(z) = \frac{1}{2} m
{\omega_z}^2 z^2$. The potential $V_2$ can in general differ from
$V_1$.  For example, in a magnetic trap the states $|1 \rangle$
and $|2 \rangle$ can have different magnetic dipole moments.
However, in the examples we consider the potentials are equal to a
good approximation, and so $V_2(z)=V_1(z)$ is assumed in the
calculations. Atom-atom interactions are characterized by the
$U_{ij} = 4 \pi N_c \hbar^2 a_{ij}/m A$, where $N_c$ is the total
number of condensate atoms and $a_{ij}$ are the s-wave scattering
lengths for binary collisions between atoms in internal states $|i
\rangle$ and $|j \rangle$.   The last pair of terms in each
equation represent the coupling via the light fields and give rise
to both coherent exchange between $|1\rangle,|2\rangle$ as well as
absorption into $|3 \rangle$. In our model, atoms which populate
$|3 \rangle$ and then spontaneously emit are assumed to be lost
from the condensate (Fig.~\ref{fig:diagram}(a)), which is why the
light coupling terms are non-Hermitian.

The light fields' propagation can be described by Maxwell's
equations.  Assuming slowly varying envelopes in time and space
(compared to optical frequencies and wavelengths, respectively)
and with the polarization densities written in terms of the BEC
wavefunctions these equations are \cite{SolitonVortex,thesis}:

\begin{eqnarray}
\label{eq:formalismPC}  \left( \frac{\partial}{\partial z}+
\frac{1}{c}\frac{\partial}{\partial t} \right) \Omega_p & = & -
\frac{N_c f_{13} \sigma_0}{2 A}(\Omega_p | \psi_1 |^2 +
\Omega_c {\psi_1}^\ast \psi_2), \nonumber \\
 \left( \frac{\partial}{\partial z}+
\frac{1}{c}\frac{\partial}{\partial t} \right) \Omega_c & = & -
\frac{N_c f_{23} \sigma_0}{2 A}(\Omega_c | \psi_2 |^2 + \Omega_p
\psi_1 {\psi_2}^\ast),
\end{eqnarray}

\noindent where $f_{13},f_{23}$ are the dimensionless oscillator
strengths of the transitions and $\sigma_0$ is the resonant cross
section ($1.65 \times 10^{-9} \mathrm{cm}^2$ and $2.91 \times
10^{-9} \mathrm{cm}^2$ for the $D_1$ lines of sodium and Rb-87,
respectively). These equations ignore quantum fluctuations of the
light fields, just as the GP equations (\ref{eq:formalism12})
ignore quantum fluctuations of the atomic fields.  Note that in
the absence of any coherence $\psi_1^* \psi_2$ these equations
predict absorption of each field with the usual two-level atom
resonant absorption coefficient (the atom density times the single
atom cross-section).

For much of our analysis, we will solve
Eqs.~(\ref{eq:formalism12})-(\ref{eq:formalismPC}) self
consistently with a numerical code, described in
\cite{SolitonVortex,thesis,numerics}. In \cite{SolitonVortex},
this was successfully applied to predict and analyze the
experimental observation of the nucleation of solitons and
vortices via the {\it light roadblock}. When the light fields are
off, the two states do not exchange population amplitude and
Eqs.~(\ref{eq:formalism12}) reduce to coupled Gross-Pitaevskii
equations for a two-component condensate. By contrast, when they
are on and the probe pulse length $\tau_0$ (and inverse Rabi
frequencies  $\Omega_p^{-1},\Omega_c^{-1}$) are much faster than
the time scale for atomic dynamics (typically $\sim$~ms), the
internal couplings induced by the light field couplings will
dominate the external dynamics.

\begin{figure}
\includegraphics{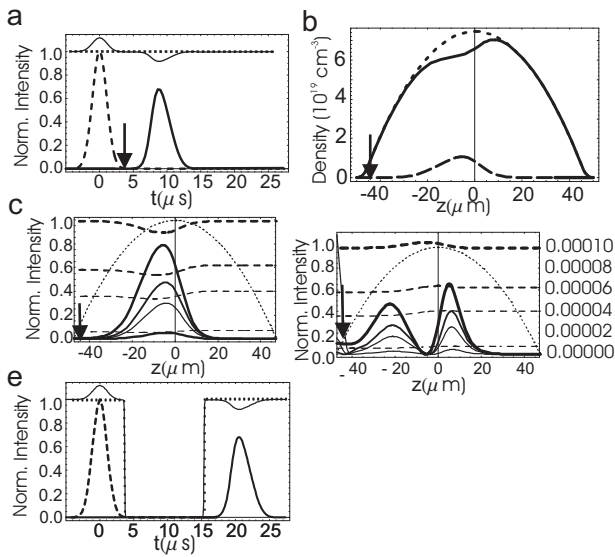} \caption{
\label{fig:switching}\textbf{Coherent storage of a light pulse via
fast switching.} Results of numerical integration of
Eqs.~(\ref{eq:formalism12})-(\ref{eq:formalismPC}) showing both a
slow and stopped light experiment.   \textbf{(a)} The dashed curve
shows the normalized {\it input} probe pulse intensity
$|\Omega_p(z_{\mathrm{in}},t)|^2/\Omega_{p0}^2$ (a
$\tau_0=1.5~\mu$s $1/e$ half-width, $\Omega_{p0}=(2 \pi)~3.5~$MHz
pulse), while the dotted curve shows
$|\Omega_c(z_{\mathrm{in}},t)|^2/\Omega_{c0}^2$ for a constant
input coupling field $\Omega_{c0}=(2 \pi)~8.0~$MHz.  The thick
solid curve shows the delayed output  probe pulse (at
$z_{\mathrm{out}}$, see Fig.~\ref{fig:diagram}(b)).  The time that
the tail of the input has vanished and the rise of the output has
not begun (marked by the arrow) corresponds to the time the pulse
is completely compressed inside the BEC. In the output coupling
intensity (thin solid curve) we see adiabatons. The values
$\sigma_0=1.65 \times 10^{-9}\,\mathrm{cm}^2$ and $f_{13}=1/2$ and
$f_{23}=1/3$ have been used.  \textbf{(b)} The atomic density in
the two states, $N_c|\psi_1|^2/A$ (solid curve) and
$N_c|\psi_2|^2/A$ (dashed curve), at $t=3.7~\mu$s (indicated by
the arrow in (a)). The dotted curve shows the original density
$N_c|\psi_1^{(\mathrm{G})}|^2/A$ before the probe is input.  The
arrow here and in (c)-(d) marks the position $z_c$ discussed later
in the text. \textbf{(c)} Spatial profiles of the probe
$|\Omega_p|^2/\Omega_{p0}^2$ (solid curves) and coupling
$|\Omega_c|^2/\Omega_{c0}^2$ (dashed curves) field intensities at
various times while the coupling field input is switched off. The
switch-off is an error function profile with a width
$\tau_s=0.1~\mu$s, centered at $t_{\mathrm{off}}=3.7\,\mu$s.
Successively thinner curves refer to $t=3.52,3.65,3.68$, and
$3.75\,\mu$s.  The dotted curve indicates the original condensate
density (arb. units). \textbf{(d)} Spatial profiles of the
normalized dark $|\Omega_D|^2/\Omega_{c0}^2$ (dashed curves, scale
on the left) and absorbed $|\Omega_A|^2/\Omega_{c0}^2$ (solid
curves, scale on right) field intensities, at the same times as in
(c). \textbf{(e)} The input and output probe and coupling
intensities, with the same conventions as (a), in a stopped light
simulation. In this case, the coupling field is switched off at
$t_{\mathrm{off}}=3.7~\mu\mathrm{s}$ and then back on at
$t_{\mathrm{on}}=15.3~\mu\mathrm{s}$.}
\end{figure}

For our our initial conditions, we consider $N_c \sim 10^6$ atoms
are initially in $|1 \rangle$ in the condensed ground state. We
determine the wavefunction of this state
$\psi_1^{(\mathrm{G})}(z)$ numerically by propagating
(\ref{eq:formalism12}) in imaginary time \cite{imaginaryTime},
though the Thomas-Fermi approximation \cite{TF} provides a good
analytic approximation to $\psi_1^{(\mathrm{G})}(z)$.  In the
cases presented in this paper, we choose the transverse area $A$
so that the central density and chemical potential $\mu=U_{11}
|\psi_1^{(\mathrm{G})}(0)|^2$ are in accordance with their values
in a trap with transverse frequencies $\omega_x=\omega_y=3.8 \,
\omega_z$, as in previous experiments \cite{SolitonVortex}.  In
Fig.~\ref{fig:switching} we consider an example with $N_c = 1.2
\times 10^6$ sodium atoms (with $a_{11}=2.75$~nm
\cite{scatteringLengthNa}), $\omega_z=(2\pi)21~\mathrm{Hz}$ and
$A=\pi(8.3~\mu\mathrm{m})^2$. The ground state density profile
$N_c |\psi_1^{(\mathrm{G})}(z)|^2$ is indicated with the dotted
curve in Fig.~\ref{fig:switching}(b).  In this case the chemical
potential $\mu=(2 \pi)~1.2~$kHz.

With our initial ground state $\psi_1^{(\mathrm{G})}(z)$
determined, consider that we initially ($t=-\infty$) input a
constant coupling field with a Rabi frequency $\Omega_{c0}$ and
then inject a Gaussian shaped probe pulse at $z_{\mathrm{in}}$
(see Fig.~\ref{fig:diagram}(b)) with a temporal half-width
$\tau_0$ and a peak Rabi frequency $\Omega_{p0}$.  We define our
times such that $t=0$ corresponds to the time the peak of the
pulse is input. The dotted and dashed curves in
Fig.~\ref{fig:switching}(a) show, respectively, a constant
coupling input $\Omega_{c0}=(2 \pi)~8\,$MHz and a weaker input
probe pulse with peak amplitude $\Omega_{p0}=(2 \pi)3.5\,$MHz and
width $\tau_0=1.5\,\mu$s.

Solving Eqs.~(\ref{eq:formalism12})-(\ref{eq:formalismPC}) reveals
that, when the conditions necessary for EIT hold, the pulse will
compress upon entering the BEC and propagate with a slow group
velocity.  As it does this, it transfers the atoms into
superpositions of $|1 \rangle$ and $|2 \rangle$ such that
\cite{thesis}:

\begin{equation}
\label{eq:DarkState} \psi_2(z,t) \approx -
\frac{\Omega_p(z,t)}{\Omega_c(z,t)}\psi_1(z,t),
\end {equation}

\noindent which is a generalization to BECs of the single atom
{\it dark state} \cite{darkState}. Thus, the probe field
$\Omega_p$ imprints its (time and space dependent) phase and
intensity pattern in the BEC wavefunctions as it propagates.  When
(\ref{eq:DarkState}) is exactly satisfied, it is easily seen that
the two light field coupling terms in each of
Eqs.~(\ref{eq:formalism12}) cancel, meaning that $|1 \rangle$ and
$|2 \rangle$ are completely decoupled from the excited state $|3
\rangle$. However, time dependence of $\Omega_p$ causes small
deviations from (\ref{eq:DarkState}) to occur, giving rise to some
light-atom interaction which is, in fact, the origin of the slow
light propagation. If one assumes the weak probe limit
($\Omega_{p0} \ll \Omega_{c0}$) and disregards terms of order
$(\Omega_{p0}/\Omega_{c0})^2$ the group velocity of the probe
pulse is \cite{HarrisHau,thesis}:

\begin{equation}
\label{eq:Vg} V_g(z)= \frac{\Omega_{c0}^2}{\Gamma}\frac{A}{N_c
f_{13} \sigma_0 |\psi_1^{(\mathrm{G})}(z)|^2}
\end{equation}

\noindent and so is proportional the coupling intensity and
inversely proportional to the atomic density $N_c
|\psi_1^{\mathrm{(G)}}(z)|^2/A$. The half-width length of the
pulse in the medium is $L_p = \tau_0 V_g$, which is a reduction
from it's free space value by a factor $V_g/c$, while its peak
amplitude does not change.  Thus only a tiny fraction of the input
probe pulse energy is in the probe field while it is propagating
in the BEC.  Most of the remaining energy coherently flows into
the coupling field and exits the BEC, as can be seen in the small
hump in the output coupling intensity (thin solid curve in
Fig.~\ref{fig:switching}(a)) during the probe input. This has been
dubbed an adiabaton \cite{adiabatons}. The magnitude of the
adiabaton is determined by the intensity of the probe pulse.

In the weak probe limit $\psi_1(z,t)$ never significantly deviates
from the ground-state wavefunction $\psi_1(z,t) \approx
\psi_1^{(\mathrm{G})}(z)$ and the coupling field is nearly
unaffected by the propagation $\Omega_c(z,t) \approx \Omega_{c0}$
(both of these results hold to
$\mathcal{O}(\Omega_{p0}^2/\Omega_{c0}^2)$). In this case
(\ref{eq:DarkState}) shows $\psi_2$ follows the probe field
$\Omega_p$ as the pulse propagates. The arrow in
Fig.~\ref{fig:switching}(a) indicates a time where the probe has
been completely input and has not yet begun to output. During this
time, the probe is fully compressed in the BEC and
Fig.~\ref{fig:switching}(b) shows the atomic densities in $|1
\rangle$ and $|2 \rangle$ at this time. The spatial region with a
non-zero density in $|2 \rangle$ corresponds to the region
occupied by the probe pulse (in accordance with
(\ref{eq:DarkState})). Equation~(\ref{eq:DarkState}) applies to
the phases as well as the amplitudes, however the phases in this
example are homogenous and not plotted.

Once the pulse has propagated through the BEC, it begins to exit
the $+z$ side.  The energy coherently flows back from the coupling
to probe field, and we see the output probe pulse (thick solid
curve). Correspondingly, we see a dip in the coupling output at
this time.  In the experiments the delay between the input and
output probe pulses seen in Fig.~\ref{fig:switching}(a) is
measured with a PMT (Fig.~\ref{fig:diagram}(b)). This delay and
the length of the atomic cloud is used in to calculate the group
velocity. The group velocity at the center of the BEC in the case
plotted is $V_g(0)=6\,$m/s.

Note that the output pulse plotted in Fig.~\ref{fig:switching}(a)
at $z_{\mathrm{out}}$ is slightly attenuated. This reduction in
transmission is due to the EIT bandwidth.  The degree to which the
adiabatic requirement $\tau_0 \gg \Gamma/\Omega_{c0}^2$ is not
satisfied, will determine the deviation the wavefunctions from the
dark state (\ref{eq:DarkState}), which leads to absorption into
$|3 \rangle$ and subsequent spontaneous emission. Quantitatively,
this reduces the probe transmission (the time integrated output
energy relative to the input energy) to \cite{thesis}:

\begin{equation}
\label{eq:TfreqWidth} T =
\frac{\int_{-\infty}^{\infty}dt\,|\Omega_p(z_{\mathrm{out}},t)|^2}
{\int_{-\infty}^{\infty}dt\,|\Omega_p(z_{\mathrm{in}},t)|^2}=
\frac{1}{\sqrt{1 + 4 D(z_{\mathrm{out}}) \left(
\frac{\Gamma}{\tau_0 \Omega_{c0}^2}\right)^2}},
\end{equation}

\noindent where $D(z)=(N_c f_{13} \sigma_0/A)
\int_{z_{\mathrm{in}}}^{z} dz'|\psi_1^{\mathrm{(G)}}(z')|^2$ is
the optical density. In the Fig.~\ref{fig:switching} example,
$D(z_{\mathrm{out}})=390$. The peak intensity of the pulse is
reduced by a factor $T^2$ while the temporal width is increased by
$T^{-1}$ (this spreading can be seen in
Fig.~\ref{fig:switching}(a)). The appearance of the large optical
density $D(z_{\mathrm{out}})$ in (\ref{eq:TfreqWidth}) represents
the cumulative effect of the pulse seeing a large number of atoms
as it passes through the BEC. To prevent severe attenuation and
spreading, we see we must use $\tau_0 >
\tau_0^{(\mathrm{min})}\equiv 2 \sqrt{D(z_{\mathrm{out}})}\,
\Gamma/\Omega_{c0}^2$, which is $1.0~\mu$s in our example.

\section{\label{sec:switching} Fast switching and storage of coherent
optical information}

We now turn our attention to the question of stopping, storing and
reviving probe pulses.  We will show here that once the probe is
contained in the BEC, the coupling field can be switched off and
on {\it faster} than the EIT adiabatic time scale without causing
absorptions or dissipation of the information
\cite{Nature2,thesis}. While we find no requirement on the time
scale for the switching, we {\it will} obtain criteria on the
length scales of $\psi_1,\psi_2$ which must be maintained to avoid
absorption events.  While previous work
\cite{Nature2,fastStopped,stoppedTheory} had addressed the fast
switching case in the weak-probe limit, here we obtain results
that are valid even when $\Omega_{p0} \sim \Omega_{c0}$
\cite{strongStopped}.  As we will see, for shorter storage times
where the external atomic dynamics play no role, these length
scale requirements can be related to the adiabatic requirements on
the input probe pulse width $\tau_0$.  The advantage of the
current analysis is that it is easily applied to analyzing probe
revivals after the atomic dynamics have completely altered the
wavefunctions $\psi_1,\psi_2$.  In such a case it is these altered
wavefunctions, rather than the input pulse width $\tau_0$, which
is relevant.

\subsection{\label{subsec:DAanalysis} Analyzing fast switching
in the dark/absorbed basis}

Consider that we switch the coupling field input off at some time
$t_{\mathrm{off}}$ with a fast time scale $\tau_s \ll \tau_0$.
Fig.~\ref{fig:switching}(c) plots the probe and coupling
intensities as a function of $z$ at various times during a
switch-off with $\tau_s=0.1~\mu\mathrm{s}$. We see the probe
intensity smoothly ramps down with the coupling field such that
their  ratio remains everywhere constant in time. Remarkably, for
reasons we discuss below, the wavefunctions $\psi_1,\psi_2$ are
{\it completely unaffected} by this switching process. Motivated
by this, we will, in the following, assume that $\psi_1,\psi_2$ do
not vary in time during this fast switching period, and later
check this assumption.

To understand this behavior, it is useful to go into a
dark/absorbed basis for the light fields, similar to that used in
\cite{normalModes}, by defining \cite{thesis}:

\begin{eqnarray}
\label{eq:DAlight} \left(\begin{array}{c} \Omega_D \\
\Omega_A
\end{array} \right)
& = &  \frac{1}{\psi_0}
\left(\begin{array}{cc} -\psi_2^* &  \psi_1^* \\
\psi_1 & \psi_2 \end{array} \right) \left(\begin{array}{c}
\Omega_p \\ \Omega_c
\end{array} \right),
\end{eqnarray}

\noindent where $\psi_0=\sqrt{|\psi_1|^2+|\psi_2|^2}$.  From
(\ref{eq:DAlight}) one sees that when the condensate is in the
dark state (\ref{eq:DarkState}), $\Omega_A=0$.  Using the notation
$\psi_i=|\psi_i| e^{i \phi_i}$ and transforming the propagation
equations (\ref{eq:formalismPC}) according to (\ref{eq:DAlight}),
one gets:

\begin{eqnarray}
\label{eq:Dprop0} \bigg(\frac{\partial}{\partial z} + i \,
\alpha_l\bigg)\Omega_D &
= & \left( -\alpha_{NA}^*+\alpha_{12}^*\right) \Omega_A \\
\label{eq:Aprop0} \left( \frac{\partial}{\partial z}+\alpha_{A}-i
\, \alpha_l\right)\Omega_A  & = & \alpha_{NA} \Omega_D,
\end{eqnarray}

\noindent where
\begin{eqnarray}
\label{eq:alphaDef}
 \alpha_A  & \equiv &  \frac{N_c  \sigma_0}{2 A}
 \left(f_{13}|\psi_1|^2 + f_{23}|\psi_2|^2 \right), \nonumber \\
  \alpha_{12} & \equiv & \frac{N_c  \sigma_0}{2 A} (f_{13} - f_{23})
\psi_1 \psi_2, \nonumber \\
 \alpha_{NA} & \equiv & \frac{1}{\psi_0^2}\bigg[\bigg(|\psi_1|\frac{d
 |\psi_2|}{dz}-|\psi_2|\frac{d|\psi_1|}{dz}\bigg) \nonumber \\
 & & \hspace{1 cm} +i\bigg(\frac{d
 \phi_2}{dz}-\frac{d\phi_1}{dz}\bigg)|\psi_1||\psi_2|\bigg],
 \nonumber \\
\alpha_l & \equiv & \frac{1}{\psi_0^2}\bigg(|\psi_1|^2\frac{d
 \phi_1}{dz}+|\psi_2|^2\frac{d \phi_2}{dz}\bigg),
\end{eqnarray}

\noindent and we have ignored the vacuum propagation terms $\sim
1/c$ in Eq.~(\ref{eq:formalismPC}). They are unimportant so long
as the fastest time scale in the problem is slow compared to time
it takes a photon travelling at $c$ to cross the condensate (about
$100~\mu\mathrm{m}/c \sim 1\,$ps).  Note that $\alpha_A$
represents the usual absorption coefficient weighted according to
the atomic density in each of $|1 \rangle$ and $|2 \rangle$.  The
terms $\alpha_{NA},\alpha_l$ arise from spatial variations in the
wavefunctions, which make the transformation (\ref{eq:DAlight})
space-dependent. The term $\alpha_{12}$ represents an additional
effect present when the light-atom coupling coefficient differs on
the two transitions ($f_{13} \not= f_{23})$ and is discussed in
detail in Section~\ref{subsec:f23}.

Consider for the moment a case with  $f_{13}=f_{23}$ (implying
$\alpha_{12}=0$) and assume a region in which $\psi_1,\psi_2$ are
homogenous (so $\alpha_{NA}=\alpha_l=0$).
Equation~(\ref{eq:Aprop0}) shows that the {\it absorbed} field
$\Omega_A$ attenuates with a length scale $\alpha_A^{-1}$, the
same as that for resonant light in a two-level atomic medium, and
less than $1\,\mu$m at the cloud center for the parameters here.
One would then get $\Omega_A \rightarrow 0$ after propagating
several of these lengths. Conversely, (\ref{eq:Dprop0}) shows the
{\it dark} light field $\Omega_D$ experiences no interaction with
the BEC and propagates without attenuation or delay.

However, spatial dependence in $\psi_1,\psi_2$ gives rise to
$\alpha_{NA},\alpha_l\not=0$ in
(\ref{eq:Dprop0})-(\ref{eq:Aprop0}), introducing some coupling
between $\Omega_D$ and $\Omega_A$, with the degree of coupling
governed by the spatial derivatives $d \psi_1/dz, \,d \psi_2/dz$.
A simple and relevant example to consider is the case of a weak
ultra-slow probe pulse input and contained in a BEC, as discussed
above (see Fig.~\ref{fig:switching}(b)). The wavefunction $\psi_2$
has a homogenous phase and an amplitude which follows the pulse
shape according to (\ref{eq:DarkState}), meaning that
$\alpha_{NA}$ scales as the inverse of the pulse's spatial length
$L_p^{-1}$.

It is important to note that this coupling is determined by
spatial variations in the {\it relative} amplitude $\psi_2/\psi_1$
and does not get any contribution from variations in the {\it
total} atomic density $\psi_0^2$.  To see this we note that if we
can write the wavefunctions as $\psi_1(z)=c_1 \psi_0(z), \,
\psi_2(z)=c_2 \psi_0(z)$, where $c_1,c_2$ are constants
independent of $z$, then $\alpha_{NA}$ and $\alpha_l$ as defined
in (\ref{eq:alphaDef}) vanish.

When variations in the relative amplitude are present, examination
of (\ref{eq:Aprop0}) reveals that when the damping is much
stronger than the coupling ($|\alpha_{NA}|,\alpha_l \ll
\alpha_{A}$), one can ignore the spatial derivative term in
analogy to an adiabatic elimination procedure.  When we do this,
(\ref{eq:Aprop0}) can be approximated by:

\begin{equation}
\label{eq:omegaA} \Omega_A \approx \frac{\alpha_{NA}}{\alpha_A}
\Omega_D
\end{equation}

\noindent Strictly speaking, one can only apply this procedure in
the region where $\Omega_A$ has propagated more than one
absorption length (that is for $z>z_c$ where $z_c$ is defined by
$\int_{z_\mathrm{in}}^{z_c} dz' \alpha_A(z')=1$). However, $z_c$
is in practice only a short distance into the BEC (marked with
arrows in Figs.~\ref{fig:switching}(b)-(d)).  When the probe has
already been input, as in Fig.2(b), $\psi_2$ and therefore
$\Omega_A$ (see (\ref{eq:DAlight})) are already trivially zero for
$z<z_c$ so (\ref{eq:omegaA}) holds everywhere. We next plug
Eq.~(\ref{eq:omegaA}) into Eq.~(\ref{eq:Dprop0}), giving:

\begin{equation}
\label{eq:Dprop} \frac{\partial \Omega_D}{\partial z} = \left[
-\frac{|\alpha_{NA}|^2}{\alpha_A} + \frac{ \alpha_{NA}
\alpha_{12}^*}{\alpha_A} - i \, \alpha_l \right] \Omega_D.
\end{equation}

\noindent These last two equations rely only on assumptions about
the spatial derivatives of $\psi_1,\psi_2$ and {\it not} on the
time scale of the switch-off $\tau_s$ (though we have assumed
$\tau_s \gg \Gamma^{-1}$ in adiabatically eliminating $|3 \rangle$
from our original
Eqs.~({\ref{eq:formalism12})-(\ref{eq:formalismPC}})).

These results allow us to conclude two important things which hold
whenever $|\alpha_{NA}|,\alpha_l \ll \alpha_A$ and the probe has
been completely input. First, the coefficients governing the
propagation of $\Omega_D$ in (\ref{eq:Dprop}) are extremely small.
The length scales $|\alpha_{NA}|^{-1},\alpha_{12}^{-1}$ are
already generally comparable to the total BEC size and the length
scales for changes in $\Omega_D$, given by (\ref{eq:Dprop}), scale
as these terms multiplied by the large ratio
$\alpha_A/\alpha_{NA}$. (The $\alpha_l$ term in (\ref{eq:Dprop})
can lead to additional phase shifts, which we discuss in
Section~\ref{subsec:f23}). Therefore the dark field $\Omega_D$
propagates with very little attenuation. As we have noted (and
later justify and discuss in more detail) the wavefunctions (and
therefore the propagation constant in brackets in
(\ref{eq:Dprop})) are virtually unchanged during the switch-off.
Under this condition, changes initiated in $\Omega_D$ at the
entering edge $z_{\mathrm{in}}$ quickly propagate across the
entire BEC. To apply this observation to the switch-off, we note
when the pulse is contained so $\psi_2=0$ at the entering edge,
(\ref{eq:DAlight}) shows $\Omega_D=\Omega_c$ there. Switching off
the coupling field at $z_{\mathrm{in}}$ then amounts to switching
off $\Omega_D$ at $z_{\mathrm{in}}$ and (\ref{eq:Dprop}) shows
that this switch-off propagates through the entire BEC with little
attenuation or delay.  Second, from (\ref{eq:omegaA}) we see that
as $\Omega_D$ is reduced to zero $\Omega_A$ is reduced such that
the ratio $\Omega_A/\Omega_D$ remains constant in time.

A numerical simulation corroborating this behavior is plotted in
Fig.~\ref{fig:switching}(d).  The dark field intensity
$|\Omega_D|^2$ is seen to switch-off everywhere as the coupling
field input is reduced to zero over a $\tau_s=0.1\,\mu$s
timescale, confirming that the changes in the input propagate
across the BEC quickly and with little attenutation. The ratio
$|\Omega_A|^2/|\Omega_D|^2$ is everywhere much smaller than unity
and constant in time. The only exception to this is in a small
region $z<z_c$ at the cloud entrance, where $\Omega_A$ has not yet
been fully damped.  The plot of $|\Omega_A|^2$ during the
switch-off indeed demonstrates how (\ref{eq:omegaA}) is a
generally good approximation, but breaks down in this region. In
the case plotted (and any case where the probe is fully contained)
the wavefunction $\psi_2$ is so negligible in this region that
$\Omega_A$ is rather small and unimportant (note the scale on the
right hand side of the plot).

Translating this back into the $\Omega_p,\Omega_c$ basis, we note
from (\ref{eq:DAlight}) that keeping $\Omega_A \approx 0$ means
the probe must (at all $z$) constantly adjust to the coupling
field via:

\begin{equation}
\label{eq:omegaPadjust}
\Omega_p=-\left(\frac{\psi_2}{\psi_1}\right)\Omega_c
\end{equation}

\noindent Thus we see that $\Omega_p$ smoothly ramps down with
$\Omega_c$ even if $\Omega_c$ is ramped down quickly, as seen in
Fig.~\ref{fig:switching}(c).

Using our results for the light fields, we can now see why the
wavefunctions $\psi_1,\psi_2$ do not change during the switching.
Physically, the probe is in fact adjusting to maintain the dark
state (\ref{eq:omegaPadjust}), and in doing so induces some
transitions between $|1 \rangle$ and $|2 \rangle$. However, only a
fraction $V_g /c \sim 10^{-7}$ of the input energy is in the probe
while it is contained in the medium, and so the probe is
completely depleted before any significant change occurs in
$\psi_1,\psi_2$ \cite{Nature2}.  In fact, the energy content of
the probe field right before the switch-off is less than 1/100th
of a free-space photon in the case here.  Thus, $\Omega_p$ is
completed depleted after only a fraction of one $|1 \rangle
\rightarrow |2 \rangle$ transition. Note that
(\ref{eq:omegaPadjust}) is equivalent to (\ref{eq:DarkState}).
However, writing it in this way emphasizes that, during the
switching, the probe is being driven by a reservoir consisting of
the coupling field and atoms and adjusts to establish the dark
state.  This is contrast to the situation during the probe input,
when many photons from both fields are being input at a specific
amplitude ratio, forcing the {\it atomic fields} to adjust to
establish the appropriate dark state. Plugging in our results for
the light fields (\ref{eq:omegaA})-(\ref{eq:Dprop}) into our
Eqs.~(\ref{eq:formalism12}) we can calculate the changes that
occur in $\psi_1$ and $\psi_2$ during the switch-off. Doing this,
we find relative changes in $\psi_1,\psi_2$ are both smaller than
$\tau_s/\tau_0$, which can be made arbitrarily small for fast
switching $\tau_s \ll \tau_0$. The little change which does occur
is due not to any process associated with the switching itself but
is due to the small amount of propagation during the switch-off.
It is therefore safe to assume (as we saw numerically) that
$\psi_1,\psi_2$ are constant in time during the switch-off. One
can also show with this analysis that the ratio of {\it coherent}
exchange events (from $|1 \rangle$ to $|2 \rangle$ or vice-versa)
to {\it absorptive} events (transitions to $| 3\rangle$ followed
by spontaneous emission) is $|\Omega_D/\Omega_A|$ implying that
the switch-off occurs primarily via coherent exchanges.

In the original stopped light experiment \cite{Nature2}, the
coupling field was then switched back on to $\Omega_{c0}$, after
some controllable storage time $\tau_{st}$, at a time
$t_{\mathrm{on}}=t_{\mathrm{off}}+\tau_{st}$.  In that experiment,
revivals were observed for storage times $\tau_{st}$ too short for
significant external atomic dynamics to occur, so the state of the
atoms was virtually identical at $t_{\mathrm{on}}$ and
$t_{\mathrm{off}}$. Then the analysis of the switch-on is
identical to the switch-off as it is just the same coherent
process in reverse.  The probe is then restored to the same
intensity and phase profile as before the switch-off. An example
of such a case is plotted in Fig.~\ref{fig:switching}(e). The
output revived pulse then looks exactly like the normal USL pulse
(compare the output in Figs.~\ref{fig:switching}(a),(e)), as was
the case in the experiment.

We have established the requirement $|\alpha_{NA}| \ll \alpha_{A}$
is necessary and sufficient for coherent switching to occur. Under
what conditions is this satisfied?  When the switch-off occurs
while the probe is compressed, it is {\it always} satisfied
because of bandwidth considerations mentioned above (see
Eq.~(\ref{eq:TfreqWidth})). Specifically, the input pulse must
satisfy $\tau_0 > \tau_0^{(\mathrm{min})}$. However, this leads to
a pulse width in the medium of $L_p > 2 \sqrt{D(z_{\mathrm{out}})}
\, \alpha_A^{-1}$, implying that $|\alpha_{NA}| \ll \alpha_{A}$ is
satisfied (as $D(z_{\mathrm{out}}) \gg 1$). Therefore, any pulse
which can successfully propagate to the cloud center, can be
abruptly stopped and coherently depleted by a rapid switch-off of
the coupling field. Of course, if the switch-on is then done
before significant atomic dynamics, the same reasoning applies
then.  In this case, our requirements on the spatial derivatives
are already encapsulated by the adiabatic requirements on the
probe pulse. It is when the external atomic dynamics during the
storage significantly change $\psi_1,\psi_2$, that the analysis of
the switch-on becomes more complicated.  This is the central
purpose of the Sections~\ref{sec:processing}-\ref{sec:systematic}
below.

As hinted above, when the pulse is not yet completely input, the
switching can cause absorptions and dissipation of the
information. In this case $\psi_2$ is significant in the region
$z<z_c$ and so $\Omega_A$ is significant for several absorption
lengths into the BEC.  Physically, the coupling field sees atoms
in a superposition of $|1 \rangle$ and $|2 \rangle$ immediately
upon entering the condensate rather than only $|1 \rangle$ atoms.
Numerical simulations of this situation confirm this. During both
the switch-on and switch-off, a significant number of atoms are
lost from both condensate components, primarily concentrated in
the $z<z_c$ region, and the revived probe pulse is significantly
attenuated relative to the pulse before the storage.

Incidentally, this explains the apparent asymmetry between the
probe and coupling in a stopped light experiment.  Whenever {\it
both} fields are being input, the temporal variations in both
fields must be slow compared with the EIT adiabatic time scale, as
the atomic fields must adjust their amplitudes to prevent
absorptions. However, when one of the fields is no longer being
input (like a contained probe pulse), the other field's input can
be quickly varied in time.

\subsection{\label{subsec:FasterSwitching} Switching faster than $\Gamma^{-1}$}

When the switching is done quickly compared to the natural
lifetime of $|3 \rangle$ ($\tau_s < \Gamma^{-1}$), the assumptions
needed for adiabatic elimination of $\psi_3$ are no longer valid.
The analysis of the switching then becomes more complicated but,
remarkably, we find the quality of the storage is not reduced in
this regime.

\begin{figure}
\includegraphics{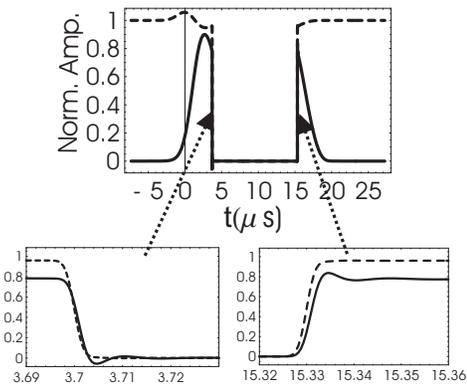} \caption{
\label{fig:fastSwitching} \textbf{Switching faster than the
natural linewidth.} Solid and dashed lines show, respectively, the
normalized probe and coupling amplitudes at a point in the center
of the BEC $\Omega_p(z=-12~\mu \mathrm{m},t)/\Omega_{p0}, \;
\Omega_c(z=-12~\mu \mathrm{m} ,t)/\Omega_{c0}$ when the switching
is faster than the natural linewidth $\tau_s=2~\mathrm{ns} \ll
\Gamma^{-1}$ (but otherwise the same parameters as
Fig.~\ref{fig:switching}(e)). The two insets magnify the regions
near the $t_{\mathrm{off}}$ and $t_{\mathrm{on}}$. }
\end{figure}

To see this, we performed numerical simulations of switching in
this regime by solving Eqs.~(\ref{eq:fullGP1})-(\ref{eq:Maxwell})
(which are the analogues of
(\ref{eq:formalism12})-(\ref{eq:formalismPC}) without the
adiabatic elimination of $\psi_3$) for the same parameters as
Fig.~\ref{fig:switching} but varying $\tau_s$ from
$80~\mathrm{ns}$ to $0.1~\mathrm{ns}$ \cite{numerics3Comp}. An
example with $\tau_s=2~\mathrm{ns}$ is shown in
Fig.~\ref{fig:fastSwitching}. We plot the probe and coupling
amplitudes versus time at a point near the center of the BEC, and
zoom in on the regions near the switch-on and -off.  Note that
that while the coupling field smoothly varies to it's new value in
each case, the probe field slightly overshoots the value given by
(\ref{eq:omegaPadjust}) before returning to it.  In all cases with
$\tau_s < \Gamma^{-1}$ the probe amplitude $|\Omega_p|$
experienced damped oscillations before reaching its final value.
The frequency of the oscillations was determined by $\tau_s^{-1}$
and the damping rate was $\Gamma$.

During these oscillations, the absorbing field $\Omega_A$ is in
fact quite significant.  However, the maximum value of $|\psi_3|$
and the total atomic loss due to spontaneous emissions events was
{\it completely independent} of $\tau_s$ over the range we
explored. To understand this, we note that even though $\Omega_A$
is larger, the frequency components of the quickly varying
coupling field are spread over a range much wider than $\Gamma$.
Thus a very small fraction of the energy in the field $\Omega_c$
is near resonant with the atomic transition and it passes through
the BEC unattenuated. Analyzing the problem in Fourier space one
can see that the increase in $\Omega_A$ for smaller $\tau_s$ is
precisely offset by this frequency width effect, leading us to the
conclusion that spontaneous emission loss is independent of
$\tau_s$.

Unlike the case of switching slowly compared with $\Gamma^{-1}$,
the adjustment of the probe is not a coherent process.  Rather,
spontaneous emission damps the system until the dark state
(\ref{eq:omegaPadjust}) is reached.  However, because we are in
the regime $V_g \ll c$ the amount of energy in the probe is much
less than one photon, and thus still has virtually no impact on
$\psi_1,\psi_2$.  This will not necessarily hold when $V_g \sim c$
and in \cite{stoppedTheory} it was predicted that there are, in
fact, adiabatic requirements for $\tau_s$ in this regime. However,
in all cases of interest here this inequality is well satisfied
and there is no restriction on $\tau_s$.

\section{\label{sec:processing} BEC dynamics and processing optical information}

We now turn to the question of the atomic dynamics during the
storage time.  These dynamics depend strongly on the relative
scattering lengths of the states used, and the probe to coupling
intensity ratio.  We will see that the these dynamics are written
onto revived probe pulses by switching the coupling field back on.
The length scale requirement on variations in the relative
amplitude $\psi_2/\psi_1$ derived above ($|\alpha_{NA}| \ll
\alpha_A$) plays a central role in the fidelity with which these
dynamics are written. We will demonstrate with several examples,
relevant to current Rb-87 experimental parameters.

\subsection{\label{subsec:general} Writing from wavefunctions onto light fields}

In \cite{Nature2} the revived pulses were seen to be attenuated
for longer storage times with a time constant of $\sim 1$~ms.  The
relative phase between the $|1 \rangle$ and $|2 \rangle$ (or
coherence), averaged over all the atoms, washed out because the
atoms, although at a cold temperature of $0.9~\mu \mathrm{K}$,
were above the Bose-condensation temperature and so occupied a
distribution of energy levels. By contrast, a zero temperature
two-component BEC will maintain a well defined phase (the phase
between $\psi_1$ and $\psi_2$) at all $z$, even as $\psi_1,\psi_2$
evolve. They evolve according to the two-component coupled GP
equations (\ref{eq:formalism12}) with the light fields set to zero
($\Omega_p=\Omega_c=0$), and the initial conditions are determined
by the superposition created by the input pulse at
$t_{\mathrm{off}}$ (see Fig.~\ref{fig:switching}(b)). After such
an evolution, we can then switch the coupling field back on and
the probe will be revived according to (\ref{eq:omegaPadjust}) but
with the {\it new} wavefunctions $\psi_1,\psi_2$, evaluated at
$t_{\mathrm{on}}$.

Here we examine a wide variety of different two-component BEC
dynamics which can occur after a pulse has been stopped. Because
of the initial spatial structure of $\psi_1,\psi_2$ created with
the USL pulses, we see some novel effects in the ensuing dynamics.

Upon the switch-on, in many cases the probe will be revived via
(\ref{eq:omegaPadjust}).  We label the spatial profile of the
revived probe by $\Omega_p^{(\mathrm{rev})}(z) \equiv
\Omega_p(z,t_{\mathrm{on}}+\tau_s)$. When we are in the weak probe
limit ($|\Omega_p| \ll |\Omega_c|$ or, stated in terms of the
wavefunctions, $|\psi_2| \ll |\psi_1|$), (\ref{eq:omegaPadjust})
becomes

\begin{equation}
\label{eq:probeRevive} \Omega_p^{(\mathrm{rev})}(z)
=-\frac{\psi_2(z,t_{\mathrm{on}})}{\psi_1^\mathrm{(G)}(z)}\Omega_{c0}.
\end{equation}

\noindent allowing us to make a correspondence between the revived
probe and $\psi_2$ if $\Omega_{c0}$ and $\psi_1^\mathrm{(G)}(z)$
are known.  The revived pulse will then propagate out of the BEC
to the PMT at $z_{\mathrm{out}}$. In the absence of any
attenuation or distortion during the propagation out, the spatial
features of the revived probe get translated into temporal
features.  Thus if we observe the output
$\Omega_p^{(\mathrm{out})}(t) \equiv \Omega_p(z_{\mathrm{out}},t)$
then we would deduce that revived pulse was

\begin{equation}
\label{eq:probeOut}
\Omega_p^{(\mathrm{rev})}(z)=
\Omega_p^{(\mathrm{out})}\big(\tau(z_{\mathrm{out}})-\tau(z)+t_{\mathrm{on}}\big)
\end{equation}

\noindent where $\tau(z_{\mathrm{out}})-\tau(z) \equiv
(\Gamma/\Omega_{c0}^2) (D(z_\mathrm{out})-D(z))$ is the time it
takes the probe pulse to travel from some point $z$ to
$z_\mathrm{out}$. Note that we are relying here on the fact that
we can switch the coupling field onto its full value of
$\Omega_{c0}$ with a time scale fast compared to the pulse delay
times $\tau(z_{\mathrm{out}})-\tau(z)$. A slow ramp up would lead
to a more complicated relationship between
$\Omega_p^{(\mathrm{rev})}(z)$ and $\Omega_p^{(\mathrm{out})}(t)$.

Combining (\ref{eq:probeRevive}) and (\ref{eq:probeOut}) shows how
the phase and amplitude information that was contained in $\psi_2$
at the time of the switch-on is transferred to the output probe
$\Omega_p^{(\mathrm{out})}(t)$. This transfer will be imperfect
for three reasons. First, when sufficiently small spatial features
are in $\psi_2$, then $\alpha_{NA}$ is comparable to $\alpha_A$,
giving rise to a significant $\Omega_A$. The resulting absorptions
will cause deviations of $\Omega_p^{(\mathrm{rev})}$ from our
expectation (\ref{eq:probeRevive}). Second, we mentioned how
spatial features are translated into temporal features on the
output probe.   During the output, fast time features on the
output probe will be attenuated via the bandwidth effect discussed
in (\ref{eq:TfreqWidth}), affecting the accuracy of the
correspondence (\ref{eq:probeOut}). Finally, stronger output
probes, which will occur when $|\psi_2| \sim |\psi_1|$ at the time
of the switch-on, make both the writing at the switch-on and the
subsequent propagation out more complicated and less reliable. Our
following examples will demonstrate these considerations.

\subsection{\label{subsec:fringes} Formation and writing of interference fringes}

Interesting dynamics occur in when the two internal states are
trapped equally ($V_2=V_1$) and the scattering lengths
$a_{12},a_{11}$ are slightly different.  We consider a case with
$N_c=1.0 \times 10^6$ Rb-87 atoms and choose $|1 \rangle =
|5S_{1/2},F=2, M_F=+1 \rangle$, $|2 \rangle = |5S_{1/2},F=1,
M_F=-1 \rangle$, and $|3 \rangle = |5P_{1/2},F=2, M_F=0 \rangle$.
The two lower states $|1 \rangle$ and $|2 \rangle$ are
magnetically trapped with nearly identical magnetic moments, and
we consider a trap with $\omega_z=(2 \pi)~21~\mathrm{Hz}$ and
assume a tranvserse area $A=\pi(5~\mu\mathrm{m})^2$. These two
states have an anomalously small inelastic collisional loss rate
\cite{inelastic} and so have been successfully used to study
interacting two-component condensates for hundreds of milliseconds
\cite{multipleComponentJILA}.  The elastic scattering lengths are
$a_{11}=5.36~\mathrm{nm}, \, a_{12}=1.024 \, a_{11}, \,
a_{22}=1.057 \, a_{11}$ \cite{scatteringLengthRb}.

\begin{figure}
\includegraphics{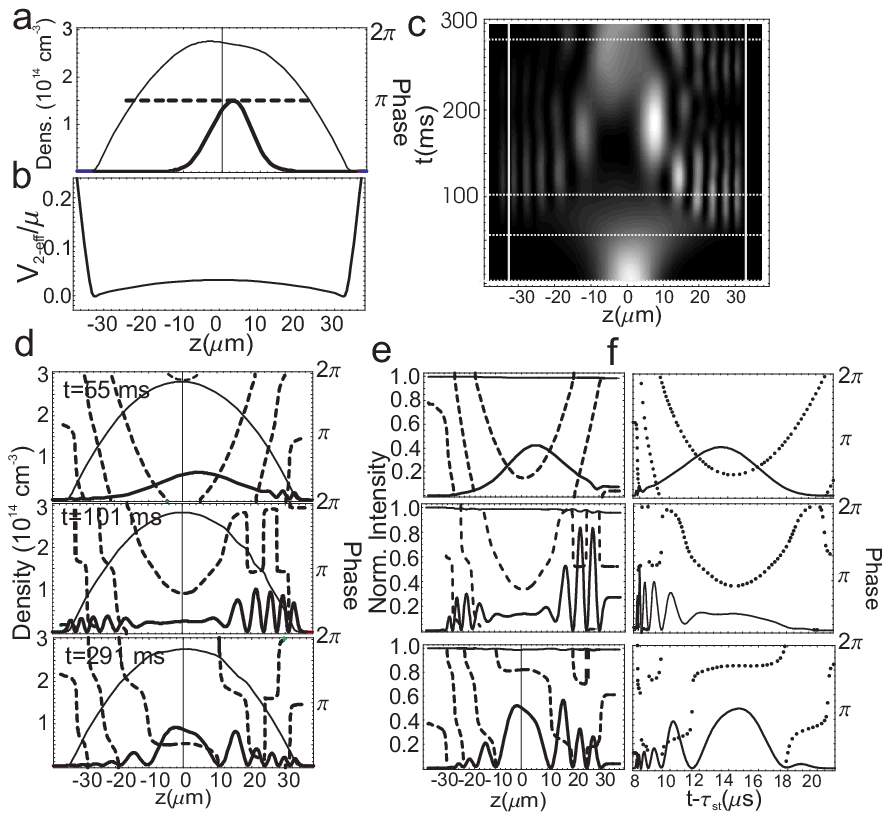}
\caption{ \label{fig:fringeEx}\textbf{Formation of interference
fringes in Rb-87 and writing fringes to probe light field.}
\textbf{(a)} The initial wavefunctions created by a $\tau=2.0~\mu
\mathrm{s}, \, \Omega_{p0}=(2 \pi)~2~\mathrm{MHz}$ probe pulse.
The coupling field is initially on at $\Omega_{c0}=(2
\pi)~8~\mathrm{MHz}$ then switched off at
$t_{\mathrm{off}}=7.5~\mu\mathrm{s}$. The cloud parameters and
atomic levels are described in the text.  The thick solid curve
shows the density $N_c |\psi_2(z)|^2/A$ (multiplied by 10 for
clarity) while the thick dashed curve shows the phase $\phi_2(z)$.
The thin curve shows the density $N_c |\psi_1(z)|^2/A$. The phase
$\phi_1$ is approximately homogenous and not shown. \textbf{(b)}
The effective potential $V_{\mathrm{2-eff}}/\mu$ (\ref{eq:V2eff})
seen by the condensate in $|2 \rangle$. \textbf{(c)} The resulting
evolution of the density $N_c|\psi_2(z,t)|^2/A$ (white represents
higher density) as the wave packet is pushed down the both sides
of potential hill and interferes with itself upon reaching the
hard wall in $V_{\mathrm{2-eff}}$ at the condensate edge. The
vertical grey lines show the condensate boundary, and the
horizontal grey lines indicate the times plotted in (d).
\textbf{(d)} Plots of the density and phase of the wavefunctions
at the indicated times (with the same conventions as (a)) at
$t=55~\mathrm{ms},101~\mathrm{ms}$ and $291~\mathrm{ms}$.
\textbf{(e)} Spatial profiles of the probe and coupling field
directly after a fast switch-on after a storage time $\tau_{st} =
t_{\mathrm{on}}-t_{\mathrm{off}}$ corresponding to the times in
(d). The thick (thin) solid curves show the intensities
$|\Omega_p^{\mathrm{(rev)}}(z)/\Omega_{p0}|^2 \,
(|\Omega_c(z)/\Omega_{c0}|^2)$ and the dashed curve shows the
phase $\phi_p^{\mathrm{(rev)}}$ ($\phi_c$ is nearly homogenous and
not plotted). \textbf{(f)} The time profile of the probe output
intensity (solid curve) and phase (dots) at the output
$z_{\mathrm{out}}$.}
\end{figure}

Figure~\ref{fig:fringeEx}(a) shows the initial wavefunctions
$\psi_1,\psi_2$ after the input and stopping of a probe pulse.  We
see a Gaussian shaped density profile in $|2 \rangle$ reflecting
the input probe's Gaussian shape.  The subsequent dynamics are
governed by Eqs.~(\ref{eq:formalism12}) with
$\Omega_p=\Omega_c=0$. In this case, a weak probe
($\Omega_{p0}^2=\Omega_{c0}^2/16$) was input.  As a result,
$\psi_1(z,t) \approx \psi_1^{(\mathrm{G})}(z)$ is nearly constant
in time and the evolution of $\psi_2$ is governed by essentially
linear dynamics, with a potential determined by the magnetic trap
{\it and} interactions with $|1\rangle$ atoms:

\begin{equation}
\label{eq:V2eff}V_{2-\mathrm{eff}}(z)=V_2(z)+U_{12}|\psi_1^{(\mathrm{G})}(z)|^2-\mu.
\end{equation}

\noindent Figure~\ref{fig:fringeEx}(b) shows this potential in
this case.  The hill in the middle has a height $\mu
[(a_{12}-a_{11})/a_{11}]$ and arises because atoms feel a stronger
repulsion from the condensate in $|1 \rangle$ when they are in $|2
\rangle$. Figures~\ref{fig:fringeEx}(c-d) show the subsequent
dynamics. One sees that the $|2 \rangle$ condensate is pushed down
both sides of the potential hill and spreads. However, once it
reaches the border of the BEC, it sees sharp walls from the trap
potential. Even for the fairly moderate scattering length
difference here, there is sufficient momentum acquired in the
descent down the hill to cause a reflection and formation of
interference fringes near the walls. The wavelength of the fringes
is determined by this momentum.

What happens if one switches the coupling field back on after
these dynamics?  Figures~\ref{fig:fringeEx}(e) shows the revived
probe pulses $\Omega_p^{(\mathrm{rev})}(z)$ upon switch-ons at the
times corresponding to Fig.~\ref{fig:fringeEx}(d).  One sees a
remarkable transfer of the sharp density and phase features of the
$|2 \rangle$ condensate onto the probe field, according to
(\ref{eq:probeRevive}).  Because we are in the weak probe regime,
the coupling field intensity only very slightly deviates from it's
input value $\Omega_{c0}$.   Figures~\ref{fig:fringeEx}(f) then
shows the output probe $\Omega_p^{(\mathrm{out})}(t)$.  The sharp
interference fringes are able to propagate out, though there is
some attenuation and washing out of the features.  Note that the
features at more positive $z$ propagate out first, leading to a
mirror image-like relationship between the spatial
Fig.~\ref{fig:fringeEx}(e) and temporal Fig.~\ref{fig:fringeEx}(f)
patterns as predicted by (\ref{eq:probeOut}).

In this particular example, we successfully output many small
features from the $\psi_2$ to the output probe field.  Here the
fringes are $\sim 3~\mu\mathrm{m}$, which is still larger than the
absorption length $\sim 0.5~\mu\mathrm{m}$, but not substantially
so, leading to some small amount of dissipation during the
switch-on and output. This gives us a sense of the ``information
capacity''. The number of absorption lengths, or optical density,
which in this case is $D(z_\mathrm{out})\approx 300$, ultimately
limits the number of features (for a given desired fidelity) which
could be successfully written and output. Note also that in the
$\tau_{st}=101~\mathrm{ms}$ case, some of the $\psi_2$ amplitude
occupies the entering region $(z<z_c)$ leading to additional
imperfections in the writing process in this region.

\subsection{\label{subsec:breathe} Breathing behavior and long storage}

\begin{figure}
\includegraphics{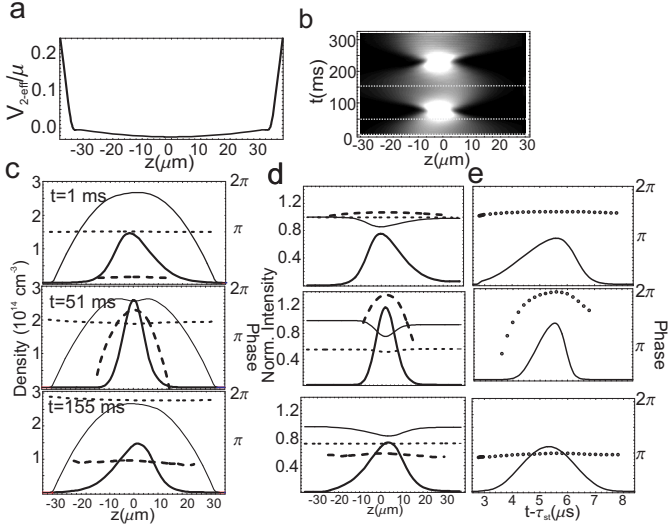} \caption{
\label{fig:breatheEx2}\textbf{Breathing behavior and long storage}
In this case the states chosen for $|1 \rangle$ and $| 2 \rangle$
are reversed, changing the sign of the curvature in the effective
potential (\ref{eq:V2eff}) in (a).  All plots are the plotted with
the same conventions as Fig.~\ref{fig:fringeEx} except we
additionally plot thin dotted curves in (c) showing the phase
$\phi_1(z)$ and in (d) showing the phase $\phi_c(z)$.}
\end{figure}

The dynamics in the previous example are quite dramatic, but are
not particularly conducive to preserving or controllably
processing the information in the BEC.  For this, it would be
preferable to switch the roles of $|1 \rangle$ and $|2 \rangle$.
Such a case is shown in Fig.~\ref{fig:breatheEx2}. Because
$a_{12}<a_{11}$ in this case, the potential hill is turned into a
trough (see Fig.~\ref{fig:breatheEx2}(a)).  The effective
potential in the region of the $|1 \rangle$ condensate, in the
Thomas-Fermi limit, becomes
$V_{2-\mathrm{eff}}=V_1[(a_{11}-a_{12})/a_{11}]$ and so it
harmonic, with a much smaller oscillator frequency than the
magnetic trap. The evolution can be easily calculated by
decomposing the wavefunction $\psi_2$ into a basis of the harmonic
oscillator states of this potential.  In the example here, there
is significant occupation of the first several oscillator levels,
and so one sees an overall relative phase shift in time (from the
ground state energy of the zeroeth state), a slight dipole
oscillation (from occupation of the first excited state) and
breathing (from the second). After one oscillator period (310~ms
in the case shown), $\psi_2$ replicates its original value at the
switch-off.

In this case, the dynamics are quite gentle, and so the spatial
scales of the phase and density features are always quite large
compared with the absorption length $\alpha_A^{-1}$. The fidelity
of the writing and output of the information on the probe field
(shown in Figs.~\ref{fig:breatheEx2}(d-e)) is correspondingly
better than in the previous example. Additionally, because the $|2
\rangle$ is trapped near the BEC center, this case avoids problems
associated with $\psi_2$ occupying the region $z<z_c$.

Because of both the ease of analyzing the evolution and the high
fidelity of outputting the information, we expect this case to be
well-suited to controlled processing of optical information. For
example, if the input pulse created a wavefunction $\psi_2$
corresponding to the ground state of the oscillator potential, the
evolution would result only in a homogenous phase shift,
proportional to the storage time, allowing long storage of the
information or introduction of controllable phase shifts.  By
choosing the pulse lengths differently so several oscillator
states are occupied, one could achieve linear processing or pulse
reshaping.

Furthermore, note that in this example there is a small but
discernable dip in the $| 1 \rangle$ density at 51~ms, indicating
some nonlinearity in the evolution of $\psi_2$. One can tune this
nonlinearity by varying the probe to coupling ratio, leading to
nonlinear processing of the information.

\subsection{\label{subsec:SP} Strong probe case: Two-component solitons}

\begin{figure}
\includegraphics{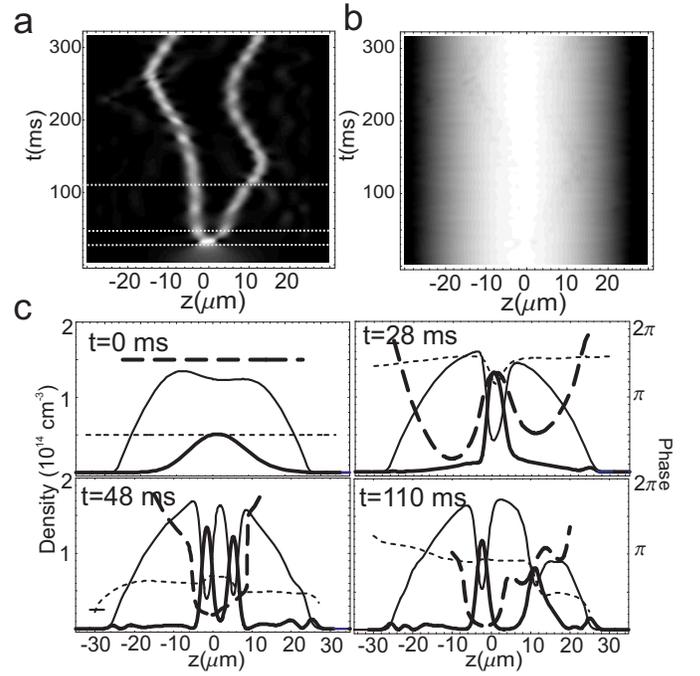} \caption{
\label{fig:solitonDynEx}\textbf{Strong probe case: Formation of
vector solitons.} Dynamics resulting from using a relatively
strong probe input $\Omega_{p0}=(2\pi)~5.7~\mathrm{MHz}, \,
\Omega_{c0}=(2\pi)~8~\mathrm{MHz}$.  In this case we have $N_c=0.5
\times 10^6$ Rb-87 atoms in a $\omega_z=(2 \pi)~21~\mathrm{Hz}$
trap and use $f_{13}=f_{23}=1/12$, and
$a_{11}=a_{22}=5.36~\mathrm{nm}$ corresponding to the system $|1
\rangle=|5S_{1/2},F=1,M_F=-1\rangle, \, |2
\rangle=|5S_{1/2},F=1,M_F=+1\rangle$, and $|3
\rangle=|5P_{1/2},F=2,M_F=0\rangle$. We artificially set
$a_{12}=1.04 \, a_{11}$ to exaggerate the phase separation
dynamics. \textbf{(a)} Evolution of density in $|2 \rangle$ $N_c
|\psi_2(z,t)|^2/A$ shows the development and interaction of two
vector solitons.  The dotted lines indicate the times plotted in
(c). \textbf{(b)} The total density $N_c
(|\psi_1(z,t)|^2+|\psi_2(z,t)|^2)/A=N_c \psi_0^2/A$ remains almost
constant in time. \textbf{(c)} The densities $N_c
|\psi_1(z,t)|^2/A$ and $N_c |\psi_2(z,t)|^2/A$ (now on the {\it
same} scale) are plotted as thin and thick solid curves,
respectively, at the times indicated.  The phases $\phi_1, \,
\phi_2$ are plotted as dotted and dashed curves.}
\end{figure}

The two-component dynamics are even richer when one uses strong
probe pulses so nonlinear effects become very evident in the
evolution. In such a case, the qualitative features of the
dynamics will be strongly effected by whether or not the relative
scattering lengths are in a phase separating regime $a_{12} >
\sqrt{a_{11} a_{22}}$ \cite{phaseSeperation,phaseSeperationTh}.
Experiments have confirmed that the $|1 \rangle, \, |2 \rangle$
studied in our previous examples in Rb-87 are very slightly in the
phase separating regime \cite{multipleComponentJILA}. Relative
scattering lengths can also be tuned via Feshbach resonances
\cite{feshbach}.

Figure~\ref{fig:solitonDynEx} shows the evolution of the two
components following the input and stopping of a stronger probe
($\Omega_{p0} = 0.71 \, \Omega_{c0}$).  In this case we chose our
levels to be $|1 \rangle=|5S_{1/2},F=1,M_F=-1\rangle, \, |2
\rangle=|5S_{1/2},F=1,M_F=+1\rangle$, and $|3
\rangle=|5P_{1/2},F=2,M_F=0\rangle$, which would require an
optical trap \cite{opticalTrap} in order to trap the $|1 \rangle$
and $|2 \rangle$ equally.  This level scheme is advantageous in
the strong probe case since $f_{13}=f_{23}$ which, as we will
discuss in more detail in Section~\ref{subsec:f23}, improves the
fidelity of the writing and output processes.  We chose the
scattering length $a_{12}=1.04 \, a_{11}$, higher than the actual
background scattering length, to exaggerate the phase separation
dynamics. One sees in Fig.~\ref{fig:solitonDynEx}(a),(c) that over
a 30~ms timescale, the phase separation causes the density in $|2
\rangle$ to become highly localized and dense. This occurs because
the scattering length $a_{12}> a_{11}$ causes $|1 \rangle$ atoms
to be repelled from the region occupied by the $|2 \rangle$ atoms,
and in turn the $|2 \rangle$ atoms find it favorable to occupy the
resulting ``well'' in the $|1 \rangle$ density. These two
processes enhance each other until they are balanced by the cost
of the kinetic energy associated with the increasingly large
spatial derivatives and we see the formation of two-component
(vector) solitons \cite{twoCompSolitons}.  In the case here, two
solitons form and propagate around the BEC, even interacting with
each other. The alternating grey and white regions along each
strip in Fig.~\ref{fig:solitonDynEx}(a) indicate that the solitons
are undergoing breathing motion on top of motion of their centers
of mass. Fig.~\ref{fig:solitonDynEx}(b) shows that the total
density profile $\psi_0^2$ varies very little in time.  It is the
{\it relative} densities of the two components that accounts for
nearly all the dynamics.

We found solitons formed even in only slightly phase separating
regimes ($a_{12}/\sqrt{a_{11} a_{22}} \ge 1.02 $).  The number of
solitons formed, the speed of their formation, and their width
were highly dependent on this ratio as well as the number of atoms
in $|2 \rangle$. The ability of stopped light pulses to create
very localized two-component structures seems to be a very
effective method for inducing the formation of vector solitons,
which has hitherto been unobserved in atomic BECs. A full
exploration of these dynamics is beyond our scope here.

\begin{figure}
\includegraphics{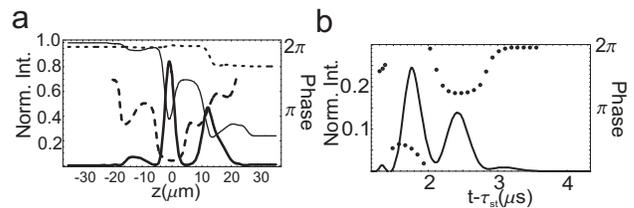} \caption{
\label{fig:solitonOutput}\textbf{Writing solitons onto probe
field.} \textbf{(a)} Writing of vector solitons onto probe and
coupling fields directly after a switch-on after a storage time of
$\tau_{\mathrm{st}}=110~\mathrm{ms}$ and \textbf{(b)} probe
output.  The plots use the same conventions as
Fig.~\ref{fig:breatheEx2}(d-e). }
\end{figure}

These non-trivial features can also be written onto probe pulses
as shown in Fig.~\ref{fig:solitonOutput}(a).  Note that the
condition for coherent revivals $|\alpha_{NA}|\ll \alpha_A$ does
{\it not} depend on the weak probe limit. Therefore, the writing
process is still primarily a coherent process, and
(\ref{eq:omegaPadjust}) is still well satisfied. However, we see
in Fig.~\ref{fig:solitonOutput}(a), the coupling field is strongly
affected by the writing and is far from homogenous, meaning the
weak probe result (\ref{eq:probeRevive}) can not be used.
Nonetheless, we see that the qualitative features of both the
density and phase of $\psi_2$ have still been transferred onto
$\Omega_p$.  The writing in the strong probe regime will be
studied in more detail in Section~\ref{sec:systematic}.

In Fig.~\ref{fig:solitonOutput}(b) we plot the resulting output
probe pulse in this case.  As in the weak probe case there is some
degradation in the subsequent propagation due to attenuation of
high frequency components.  In addition, unlike the weak probe
case, there are nonlinearities in the pulse propagation itself
which causes some additional distortion. Section~\ref{subsec:f23}
also addresses this issue. Even so, we again see a very clear
signature of the two solitons in the both the intensity and phase
of the output probe.

We have performed some preliminary simulations in traps with
weaker transverse confinement and which calculate the evolution in
the transverse degrees of freedom.  They show that the solitons
can break up into two-component vortex patterns via the snake
instability \cite{snake,SolitonVortex}.

We chose our parameters to be in the phase separating regime here.
Just as in the weak probe case, other regimes will lead instead to
much gentler dynamics.  For example, it has been shown
\cite{breatheTogether} that when $a_{12}<a_{11}=a_{22}$ then there
exist ``breathe-together'' solutions of the BEC whereby complete
overlap of the wavefunctions $\psi_1,\psi_2$ persists. In this
case the nonlinear atomic interaction can lead to spin-squeezing
\cite{spinSqueezing}. It would be extremely interesting to
investigate a probe-revival experiment in such a case, to see to
what extent the squeezed statistics are written into probe pulse,
producing squeezed light \cite{squeezedLight}. This would require
taking both the light propagation and atomic dynamics in our
formalism beyond the mean field.  Steps in this direction have
been taken in \cite{quantumProcessing,squeezedLightRaman}, but
these analyses are restricted to the weak probe case.

\section{\label{sec:systematic} Quantitative study of writing and output}

The above examples demonstrate that a rich variety of two
component BEC dynamics can occur, depending on the relative
interaction strengths and densities of the states involved. They
also show that remarkably complicated spatial features in both the
density and phase can be written into temporal features of a probe
light pulse and output.  The purpose of this section is to
quantify the fidelity with which wavefunctions can be written onto
the probe field.

To do this we consider an example of a two-component BEC with a
Gaussian shaped feature in $\psi_2$, with parameters
characterizing length scales and amplitudes of density and phase
variations. Note these Gaussian pulse shapes are relevant to cases
in which one might perform controlled processing, {\it e.g.}
Fig.~\ref{fig:breatheEx2}.   Switching on a coupling field then
generates and outputs a probe pulse with these density and phase
features. We will calculate this output, varying the parameters
over a wide range. In each case, we will compare the output to
what one expect from an ``ideal'' output (without dissipation or
distortion), and calculate an error which characterizes how much
they differ. Using the analysis of the switching process in
Section~\ref{sec:switching} and the USL propagation in
Section~\ref{sec:formalism} we also obtain analytic estimates of
this error.

In the weak probe case, we find a simple relationship between
$\psi_2(z,t_\mathrm{on})$ and the output
$\Omega_p^{(\mathrm{out})}(t)$.  We find that to optimize the
fidelity in this case one should choose the Gaussian pulse length
between two important length scales.  The shorter length scale is
determined by EIT bandwidth considerations, which primarily
contribute error during the output. On the other hand, for large
pulses comparable to the total condensate size, the error during
the writing process dominates due to $\psi_2$ partially occupying
the condensate entering edge region ($z<z_c$).

In the strong probe case, we find that the relationship between
$\psi_2(z,t_\mathrm{on})$ and $\Omega_p^{(\mathrm{out})}(t)$ is
more complicated, however, when $f_{13}=f_{23}$, one can still
make an accurate correspondence.  We find the fidelity in this
case depends only weakly on the probe strength.  Conversely, when
$f_{13}\not= f_{23}$, even a small nonlinearity causes additional
phase shifts and distortions, making this correspondence much more
difficult, leading to higher errors for stronger probes. For this
reason, a system that uses, for example $|1 \rangle = |F=1, M_F=-1
\rangle, \, |2 \rangle = |F=1, M_F=+1 \rangle$ (as in
Fig.~\ref{fig:solitonOutput}) would be preferable when one is
interested in applications where there are comparable densities in
the two states.

By considering the switch-on and output processes, starting with
an arbitrary initial two-component BEC, we emphasize that this
method of outputting the atomic field information onto light
fields works regardless of how the BEC state was generated. One
could prepare a BEC in a coherent superposition of $|1 \rangle$
and $|2 \rangle$ by any available method including, but not
limited to, inputting a slow light pulse.

\subsection{\label{subsec:fidelity} Deriving $\psi_2$ from the written and output probe field.}

Assume we have a two component BEC with wavefunctions
$\psi_1,\psi_2$ (and again define $\psi_0 \equiv
\sqrt{|\psi_1|^2+|\psi_2|^2}$).  For now we assume $f_{13}=f_{23}$
(so $\alpha_{12} = 0$, see Eq.~(\ref{eq:alphaDef})), as we treat
the $f_{13}\not=f_{23}$ case in Section~\ref{subsec:f23}.  Upon a
rapid switch-on of a coupling field with amplitude $\Omega_{c0}$,
inverting (\ref{eq:DAlight}) shows the pattern written onto the
probe field will be:

\begin{equation}
\label{eq:probeReviveEx} \Omega_p^{\mathrm{(rev)}}= -
\frac{\psi_2}{\psi_0} \Omega_D + \frac{\psi_1^*}{\psi_0} \Omega_A
\end{equation}

\noindent In the ideal limit, where the spatial variations of the
wavefunction are sufficiently small, $\alpha_{NA} \rightarrow 0$
and the second term vanishes.  Furthermore, (\ref{eq:Dprop}) shows
in this case that the dark field intensity is constant $|\Omega_D|
= \Omega_{c0}$.  So in the ideal limit we have for the amplitude:

\begin{equation}
\label{eq:probeReviveMag} |\Omega_p^{({\mathrm{rev-ideal}})}| =
\frac{|\psi_2|}{\psi_0}\Omega_{c0}
\end{equation}

\noindent In practice this will be a good approximation so long as
the inequality $|\alpha_{NA}| |\psi_1| \ll \alpha_A |\psi_2|$ is
satisfied.  The extent to which the second term in
(\ref{eq:probeReviveEx}) cannot be neglected will determine the
error between the ideal $\Omega_p^{({\mathrm{rev-ideal}})}$ and
actual $\Omega_p^{({\mathrm{rev}})}$ output.  Note that in all the
cases studied in Section~\ref{sec:processing}, $\psi_0(z,t)$ was
nearly constant in time, being always well approximated by the
original ground state $\psi_1^{\mathrm{(G)}}(z)$. Thus $\psi_0$
can be considered a known function of $z$.

Turning now to the phase, the simple relationship
$\phi_2-\phi_1=\phi_p-\phi_c+\pi$ is always satisfied in the dark
state (see (\ref{eq:omegaPadjust})). If $\phi_c$ was just a
constant then the probe phase would simply reflect the relative
phase of the two wavefunctions $\phi_p=\phi_2-\phi_1+\pi$
(choosing our phase conventions so $\phi_c=0$). This is indeed the
case in the weak probe limit, as then the coupling field phase
$\phi_c$ is unaffected by the atomic fields upon the switch-on.
However, this is not necessarily so in the strong probe regime. To
calculate the true phase shift, we calculate phase of $\Omega_D$
from (\ref{eq:Dprop}).  In the ideal limit
$\phi_D(z)=\int_{z_\mathrm{in}}^{z} dz' \alpha_l(z')$ (choosing
our phase conventions such that $\phi_1(z_\mathrm{in})=0$).
Defining $\phi_{21}\equiv \phi_2 - \phi_1$ and using the
definition of $\alpha_l$ (\ref{eq:alphaDef}) and
(\ref{eq:probeReviveMag}), we find the relationship:

\begin{eqnarray}
\label{eq:probeRevivePhase} \phi_{21}(z) & =&
\phi_p^{({\mathrm{rev-ideal}})}(z)+ \pi
-\phi_{p}^{(\mathrm{nl})}; \nonumber \\
\phi_{p}^{(\mathrm{nl})} & = & \int_{z_{\mathrm{in}}}^z dz'\frac{d
\phi_p^{({\mathrm{rev-ideal}})}}{dz'}
\frac{|\Omega_p^{({\mathrm{rev-ideal}})}(z')|^2}
{\Omega_{c0}^2-|\Omega_p^{({\mathrm{rev-ideal}})}(z')|^2}, \nonumber \\
\end{eqnarray}

\noindent The equation is written in this way (the atomic phase in
terms of the probe phase and intensity) because our purpose is to
derive the phase pattern $\phi_{21}$ based on the observed probe
field, rather than vice versa.  The nonlinear correction
$\phi_{p}^{(\mathrm{nl})}$ is fact the phase imprinted on the
coupling field $\phi_c$ during the switch-on.

However, in an experiment, we do not have direct access to the
revived probe $\Omega_p^{\mathrm{(rev)}}(z)$.  Rather, we observe
the temporal output $\Omega_p^{\mathrm{(out)}}(t)$.  In practice,
the intensity of this quantity can be measured with a PMT, while
the phase pattern could be measured by beating it with a reference
probe field which did not propagate though the BEC.  In the ideal
limit (the absence of attenuation or distortion) the relationship
(\ref{eq:probeOut}) will link the observed output to the revived
probe field $\Omega_p^{\mathrm{(rev)}}$.  Thus we define the
``ideal'' output via

\begin{eqnarray}
\label{eq:probeOutIdeal} \Omega_p^{(\mathrm{rev-ideal})}(z) \equiv
\Omega_p^{(\mathrm{out-ideal})}\big(\tau(z_{\mathrm{out}})-\tau(z)+t_{\mathrm{on}}\big)
\end{eqnarray}

\noindent In practice, additional absorption events and distortion
can occur during the output. If variations of the relative
amplitude $\psi_2/\psi_1$ have some characteristic scale $L_\psi$
then time features with a scale $\tau \sim L_\psi/V_g$ will be
introduced into the probe $ \Omega_p$. Using our result for the
bandwidth induced attenuation (\ref{eq:TfreqWidth}) yields an
estimate for transmission energy (that is, the energy of the
actual output pulse $\Omega_p^{(\mathrm{out})}$ relative to an
unattenuated output of $\Omega_p^{(\mathrm{rev})}$):

\begin{eqnarray}
\label{eq:Tspace}  T & = & \frac{1}{\sqrt{1+\beta}}; \nonumber \\
\beta & \equiv & \frac{
D(z_{\mathrm{out}})-D(z_{\mathrm{p}})}{\big(L_{\psi}
\alpha_A(z_{\mathrm{p}})\big)^2}
\end{eqnarray}

\noindent For this simple estimate we evaluate $\alpha_A$ and $D$
at the location of the center of the pulse $z_p$, though one could
also construct more sophisticated estimates by integrating over
the spatial distribution of the pulse.  Just as in our discussion
below Eq.~(\ref{eq:TfreqWidth}), the temporal width is increased
by a factor $T^{-1}$ and the peak intensity reduced by $T^2$
during the propagation out.

We now apply these findings to a Gaussian pulse, with $\psi_2$
assumed to be of the form,

\begin{equation} \label{eq:psi2test} \psi_2(z) = \psi_0(z) A_2 \,
\mathrm{exp}\bigg(-\frac{z^2}{2 L_2^2}\bigg)\mathrm{exp}\bigg[i
\frac{A_{\phi_2}}{2}\mathrm{erf}\bigg(\frac{z}{L_{\phi_2}}\bigg)\bigg]
\end{equation}

\begin{figure}
\includegraphics{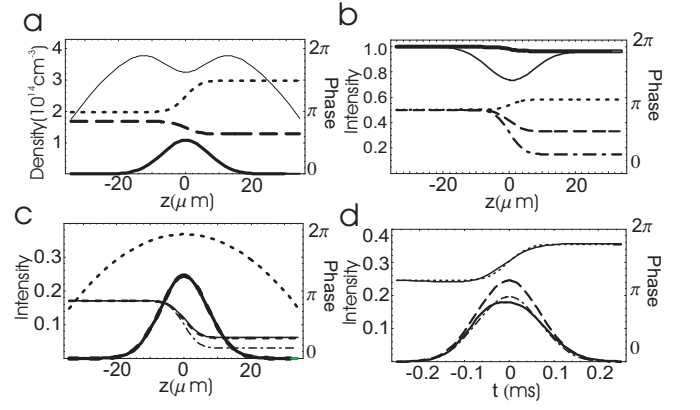}
\caption{\label{fig:test1}\textbf{Writing and outputting atomic
field information onto the probe fields.} \textbf{(a)} A test case
with a density feature of length $L_2= 10~\mu\mathrm{m}$ of
amplitude $A_2=0.5$ and phase features of $A_{\phi_2}=-0.2 \, \pi,
\, L_{\phi_2}=5~\mu \mathrm{m}$ (see Eq.~(\ref{eq:psi2test})) and
$A_{\phi_1}=0.5 \, \pi, \, L_{\phi_1}=5~\mu \mathrm{m}$.  The
conventions for the curves are the same as
Fig.~\ref{fig:solitonDynEx}(c). \textbf{(b)} Upon switch-on of the
coupling field with $\Omega_{c0}=(2 \pi)~8~$MHz, we show the dark
intensity $|\Omega_D(z)/\Omega_{c0}|^2$ (thick solid curve) and
coupling intensity $|\Omega_c(z)/\Omega_{c0}|^2$ (thin solid). The
dotted, dashed, and dot-dashed curves show, respectively, the
phase profiles $\phi_c$, $\phi_D$ and $\phi_{21}$. \textbf{(c)}
The normalized probe intensity profile
$|\Omega_p^{\mathrm{(rev)}}|^2/\Omega_{c0}^2$ (solid curve) is
indistinguishable from the expected profile
$|\Omega_p^{\mathrm{(rev-ideal)}}|^2/\Omega_{c0}^2$ (thick dashed)
(see Eq.~(\ref{eq:probeReviveMag})). The thin solid curve shows
$\phi_p^{\mathrm{(rev)}}$ and the dot-dashed curve shows
$\phi_{21}+\pi$, while the thin dashed curve shows the expected
probe phase profile with the nonlinear correction
$\phi_{p}^{\mathrm{(rev-ideal)}}=\phi_{21}+\phi_{p}^{\mathrm{(nl)}}+\pi$
(\ref{eq:probeRevivePhase}). For reference, the dotted curve shows
the total density $\psi_0^2$ profile. \textbf{(d)} The temporal
pattern of the output probe intensity
$|\Omega_p^{\mathrm{(out)}}|^2/\Omega_{c0}^2$ (thick solid curve)
and phase $\phi_p^{\mathrm{(out)}}$ (thin solid), versus an
estimate based on perfect propagation of the revived pulse
$\Omega_p^{\mathrm{(rev)}}$ according to (\ref{eq:probeOut})
(dashed and dotted curves).  The dot-dashed curve shows the ideal
output attenuated by $T^2$ (see (\ref{eq:Tspace}) and the
discussion afterward) using the estimate
$L_{\psi}=(L_2^{-2}+A_{\phi_2} L_{\phi_2}^{-2}+A_{\phi_1}
L_{\phi_1}^{-2})^{-1/2}$. }
\end{figure}

\noindent  Note $\psi_2$ has a density feature of amplitude
$A_2^2$ and length $L_2$ and a phase feature of amplitude
$A_{\phi_2}$ and length $L_{\phi_2}$.  For $\psi_1(z)$ we choose
the amplitude so the total density matches the ground state for
the trap $\psi_0^2=|\psi_1^{(\mathrm{G})}|^2$ and in the phase we
put in a shift with some amplitude $A_{\phi_1}$ and length
$L_{\phi_1}$.  An example is shown in Fig.~\ref{fig:test1}(a).
These parameters will be varied throughout this section to learn
how they effect the writing and output.  We choose a $\omega_z=(2
\pi)~20~$Hz trap ($A=\pi (5~\mu\mathrm{m})^2$) with $N_c=2.0
\times 10^6$ Rb-87 atoms.  We use $f_{13}=f_{23}=1/12$,
corresponding to the system $|1
\rangle=|5S_{1/2},F=1,M_F=-1\rangle, \, |2
\rangle=|5S_{1/2},F=1,M_F=+1\rangle$, and $|3
\rangle=|5P_{1/2},F=2,M_F=0\rangle$ in Rb-87.

In the example of Fig.~\ref{fig:test1}(a), $A_2=0.5$ is not
particularly small, so the weak probe limit can not be assumed.
Figure~\ref{fig:test1}(b) shows the spatial profiles of the dark
field $\Omega_D$ and coupling field $\Omega_c$ immediately
following a fast switch-on ($\tau_s=0.1~\mu\mathrm{s}$).  We see
very little attenuation of $\Omega_D$ occurs across the BEC, as
$\alpha_{NA}\sim \mathcal{O}(A_2 L_2,i A_2 A_{\phi_2} L_{\phi_2},i
A_2 A_{\phi_1} L_{\phi_1})$ is quite small compared with
$\alpha_{A}$. Translating this back into the $\Omega_c,\Omega_p$
basis, (\ref{eq:DAlight}) predicts that $\Omega_c$ acquires a dip
in intensity with a height proportional to the density
$|\psi_2|^2$. This behavior is indeed seen in the figure.

The phase difference $\phi_{21}$ is plotted as the dot-dashed
curve in Figure~\ref{fig:test1}(b).  One sees that in the region
where $\phi_{21}$ is inhomogeneous, a small phase shift, equal to
$\phi_{p}^{(\mathrm{nl})}$ (\ref{eq:probeRevivePhase}), is
introduced in the coupling field (dotted curve). Again this shift
only arises in the strong probe regime. The phase shift in the
dark field $\phi_D$ is plotted as the dashed curve.

Figure~\ref{fig:test1}(c) then compares written probe field to our
ideal limit predictions
(\ref{eq:probeReviveMag})-(\ref{eq:probeRevivePhase}) in this
example. The numerically calculated intensity
$|\Omega_p^{({\mathrm{rev}})}|^2$ (solid curve) is almost
indistinguishable from our prediction
$|\Omega_p^{({\mathrm{rev-ideal}})}|^2$ (\ref{eq:probeReviveMag})
(dashed curve).  The phase written onto the probe field
$\phi_p^{({\mathrm{rev}})}$ is also very close to the ideal limit
prediction (\ref{eq:probeRevivePhase}). One sees that including
the nonlinear correction $\phi_p^{\mathrm{(nl)}}$ to the simpler
estimate $\phi_{21}+\pi$ (dot-dashed curve) is important in making
the comparison good.

Figure~\ref{fig:test1}(d) then shows intensity (thick solid curve)
and phase (thin solid) of the output pulse
$\Omega_p^{({\mathrm{out}})}(t)$.  For comparison the dashed and
dotted curves show, respectively, the output expected from an
unattenuated and undistorted transmission of the revived pulse
$\Omega_p^{({\mathrm{rev}})}(z)$ via (\ref{eq:probeOutIdeal}). One
sees a very good agreement in the phase pattern, while there is a
visible reduction in the intensity, due to bandwidth
considerations.  The dot-dashed curve shows the our estimate with
the estimated reduction in intensity $T^2$ (see (\ref{eq:Tspace}))
using a characteristic length scale $L_\psi$ calculated from a
quadrature sum of contributions from the amplitude and phase
features in (\ref{eq:psi2test}) (the expression is given in the
caption).

We have thus calculated a method by which the output probe pulse
can be solely used to calculate the relative density and phase of
the wavefunctions which generated it, and demonstrated the method
with a generic example. Furthermore, we have identified the
leading order terms which will cause errors in these predictions.
In particular we have seen in our example in
Fig.~\ref{fig:test1}(d) that including the expected bandwidth
attenuation accounts for most of the deviation between our ideal
predictions and the actual output pulse.

\subsection{\label{subsec:error} Quantifying and estimating the fidelity.}

We now quantify the deviations from our predictions
(\ref{eq:probeReviveMag}-\ref{eq:probeOutIdeal}) for our example
(\ref{eq:psi2test}), varying the length and amplitude parameters
over a wide range.  These results can be directly applied to
pulses which are approximately Gaussian (as in
Fig.~\ref{fig:breatheEx2}).  The results here should also provide
a good guide to the expected fidelity in more complicated cases so
long as the length scale and amplitude of features can be
reasonably estimated.

To quantify the deviation from our ideal case prediction
(\ref{eq:probeReviveMag})-(\ref{eq:probeRevivePhase}) we define
the write error:

\begin{equation}
\label{eq:eWrite} E_w \equiv
\frac{\int_{z_{\mathrm{in}}}^{z_{\mathrm{out}}} \, dz'
\bigg|\,\Omega_p^{(\mathrm{rev})} -
|\Omega_p^{(\mathrm{rev-ideal})}|
e^{i(\phi_{21}+\phi_{p}^{\mathrm{(nl)}}+\pi)}\bigg|^2}
{\int_{z_{\mathrm{in}}}^{z_{\mathrm{out}}} \, dz'
|\Omega_p^{(\mathrm{rev})}|^2}
\end{equation}

\begin{figure}
\includegraphics{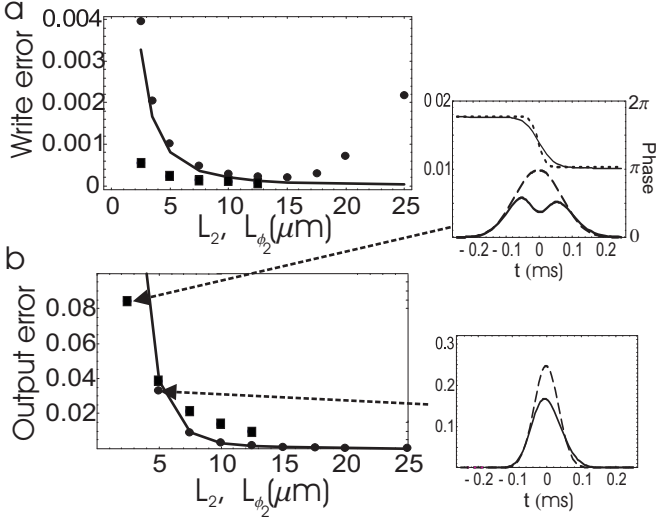}
\caption{\label{fig:seriesLength}\textbf{Writing and outputting
atomic field information onto light fields.} \textbf{(a)} Circles
show the write error $E_w$ (\ref{eq:eWrite}) versus the length
$L_2$ for BECs with no phase jumps ($A_{\phi_2}=A_{\phi_1}=0$) and
with $A_2=0.5$ (see Eq.~(\ref{eq:psi2test})). The solid curve
shows an analytic result based on the analytic expressions
(\ref{eq:probeReviveEx}), (\ref{eq:omegaA}) and (\ref{eq:Dprop}).
The squares show a case holding $L_2=10~\mu\mathrm{m},A_2=0.1$ and
introducing a phase feature $A_{\phi_2}=0.75 \, \pi$ with varying
length scale $L_{\phi_2}$. \textbf{(b)} The output error
$E_{\mathrm{out}}$ (\ref{eq:eOut}) for the same cases as (a). The
solid curves showing the prediction $(5/16)\beta^2$ discussed in
the text (choosing $L_\psi=L_2$) for the case with no phase jumps
(circles). The insets show the output in the cases indicated with
arrows, with the same plotting conventions as
Fig.~\ref{fig:test1}(d) (in the lower inset on the right $\phi_p$
is homogenous and not plotted), we again use $\omega_z=(2
\pi)~20~$Hz, $A=\pi (5~\mu\mathrm{m})^2$, $N_c=2.0 \times 10^6$
and $f_{13}=f_{23}=1/12$, corresponding to the system $|1
\rangle=|5S_{1/2},F=1,M_F=-1\rangle, \, |2
\rangle=|5S_{1/2},F=1,M_F=+1\rangle$, and $|3
\rangle=|5P_{1/2},F=2,M_F=0\rangle$ in Rb-87. }
\end{figure}

\noindent We plot this quantity in a series of cases with
different $L_2$ in Fig.~\ref{fig:seriesLength}(a) (circles) in a
case with amplitude $A_2=0.5$ and no phase profiles. This is
compared with a calculated prediction for $E_w$ (solid curve)
based on (\ref{eq:probeReviveEx}) where we calculate $\Omega_A$
with (\ref{eq:omegaA}) and calculate the small attenuation of
$\Omega_D$ with (\ref{eq:Dprop}). These errors grow as $L_2$
becomes comparable to $\alpha_A^{-1}=0.2~\mu \mathrm{m}$,
according to the discussion in Section~\ref{subsec:DAanalysis}. We
see the agreement between the analytic and numerical estimates is
quite good for small to moderate $L_2$, confirming that this is
the leading source of error.  However, when $L_2$ becomes
comparable to the total BEC size (the Thomas-Fermi radius is
$R_z=44~\mu\mathrm{m}$ here), we begin to see additional errors
because $\psi_2$ becomes non-zero at the BEC edge. Thus, we see
that for a given BEC length and density, one must choose
$\alpha_A^{-1} \ll L_2 \ll R_z$ to minimize the write error $E_w$.
Because this edge effect depends on the density structure near the
BEC edge, it is difficult to estimate analytically, but it is
related to the fraction of the pulse in or near the region
$z<z_c$.  For pulses near the BEC center, this will depend on
$L_2/R_z$, whereas for pulses far off center, it will also depend
on the location of the center of the pulse.

We also plot a series of cases keeping $L_2=10~\mu\mathrm{m}$ and
$A_2=0.1$ but varying the length scale $L_{\phi_2}$ of a phase
profile of amplitude $A_{\phi_2}=0.75 \, \pi$.  One sees similar
but slightly smaller errors $E_w$ in this case.

We define the error accumulated as $\Omega_p^{\mathrm{(rev)}}$
propagates out using (\ref{eq:probeOut}):

\begin{equation}
\label{eq:eOut} E_{\mathrm{out}} \equiv
\frac{\int_{z_{\mathrm{in}}}^{z_{\mathrm{out}}} \, dz'
|\Omega_p^{(\mathrm{rev})}(z')-\Omega_p^{(\mathrm{out})}\big(\tau(z_{\mathrm{out}})-\tau(z')\big)
|^2}{\int_{z_{\mathrm{in}}}^{z_{\mathrm{out}}} \, dz'
|\Omega_p^{\mathrm{(rev)}}(z')|^2}
\end{equation}

\noindent In Fig.~\ref{fig:seriesLength}(b) we plot this quantity
for the cases corresponding to Fig.~\ref{fig:seriesLength}(a). For
comparison, for the series with no phase shift, we calculated an
estimate based on the expected attenuation $T^2$ and spreading
$T^{-1}$ of a Gaussian pulse, due bandwidth considerations (see
Eq.~(\ref{eq:Tspace}) and subsequent discussion), choosing
$L_\psi=L_2$. In the limit of small $\beta$ this calculation
yields $E_{\mathrm{out}} \approx (5/16) \beta^2$. This is plotted
as a solid curve and we see good agreement with the numerical
data.

The error $E_\mathrm{out}$ is seen to dominate $E_w$ for small
$L_2$, due to the fact that the large optical density effects the
former ($D(z_{\mathrm{out}})=617$ in the case plotted). Thus our
analytic estimate $E_{\mathrm{out}}=(5/16)\beta^2$ is a good
estimate of the total error.  But for larger $L_2$ the edge effect
in $E_w$ becomes important. To minimize the total error
$E_w+E_\mathrm{out}$ we should choose $L_2$ so the error
$E_\mathrm{out}=(5/16) \beta^2$ is comparable to the edge effect
error. In Fig.~\ref{fig:seriesLength} the optimal length is
$L_2^{\mathrm{(opt)}} \approx 17.5~\mu \mathrm{m}$ and the total
error is $E_\mathrm{out}+E_w \approx 0.0009$.  The scaling of
$L_2^{\mathrm{(opt)}}$ with the condensate size $R_z$ is difficult
to estimate.  Assuming it roughly increases as
$L_2^{\mathrm{(opt)}} \propto R_z$, then (\ref{eq:Tspace}) shows
$\beta^{\mathrm{(opt)}} \propto 1/D(z_\mathrm{out})$, giving us a
guide as to the improvement in fidelity we can expect by
increasing the total optical density of the condensate.

The insets show the plots of the ideal output of
$\Omega_p^{\mathrm{(rev)}}(z)$ (solid curves) versus actual
outputs $\Omega_p^{\mathrm{(out)}}(t)$ (dotted curves) in two of
the cases. In one case, with no phase jumps and a small $L_2$, we
see an overall reduction in amplitude and slight spreading.  In
the other, with a larger $L_2$ but a phase shift with a small
length scale $L_{\phi_2}$, we see that the attenuation is
localized in the middle of the pulse. This is because it is the
components which contain the sharpest phase profile, near the
middle, which are most severely attenuated. Generally, in cases
with complicated spatial features, one must be aware of this
potential for local attenuation of the sharpest features.  Note,
however, that in many cases of interest, such as the interference
fringes (Fig.~\ref{fig:fringeEx}) and solitons
(Fig.~\ref{fig:solitonDynEx}), large phase shifts occur primarily
in regions of low density, meaning these features often survive
during the output propagation.

\begin{figure}
\includegraphics{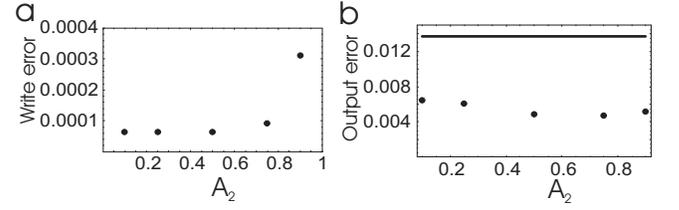}
\caption{\label{fig:seriesAmp}\textbf{Writing and outputting
atomic field information onto light fields.} \textbf{(a)} The
write error $E_w$ (\ref{eq:eWrite}) versus $A_2$ for the case
$L_2=10~\mu\mathrm{m}, \,  L_{\phi_1}=L_{\phi_2}=5~\mu\mathrm{m},
\, A_{\phi_1}=0.5 \pi, \, A_{\phi_2}=0.2 \pi$. \textbf{(b)} The
output error $E_{\mathrm{out}}$ (\ref{eq:eOut}) in the same cases.
The solid line shows the estimate $(5/16)\beta^2$ (see
Eq.~(\ref{eq:Tspace})) using $L_{\psi}=(L_2^{-2}+A_{\phi_2}
L_{\phi_2}^{-2}+A_{\phi_1} L_{\phi_1}^{-2})^{-1/2}$.  The
condensate parameters are the same as in
Fig.~\ref{fig:seriesLength}.}
\end{figure}

Our analysis of switching process in Section~\ref{sec:switching}
applies even with a large amplitude $A_2$, as was seen in the
$A_2=0.5$ case in Fig.~\ref{fig:test1}. In
Fig.~\ref{fig:seriesAmp}(a) we plot the results of a series of
simulations with $A_2$ varying all the way up to $A=0.9$ and see
only a very small impact on $E_w$ for $A_2 \le 0.8$.  Most of our
results for the propagation of USL pulses, {\it e.g.}
(\ref{eq:Vg}) and (\ref{eq:TfreqWidth}), rely on the weak probe
limit. Surprisingly though Fig.~\ref{fig:seriesAmp}(b) shows that
the output error $E_{\mathrm{out}}$ is independent of $A_2$
through $A_2=0.9$.

\subsection{\label{subsec:f23} Effect of unequal oscillator strengths}

The lack of distortion in the strong probe cases discussed so far
in this section is a result of us choosing a system with equal
oscillator strengths $f_{13}=f_{23}$. The importance of the
relative size of the oscillator strengths can be understood by
considering the adiabatons (the changes in the output coupling
intensity seen in Fig.~\ref{fig:switching}(a,e)). The adiabatons
arise because of the coherent flow of photons between the two
light fields and their amplitude is determined by the requirement
that the sum of the number of photons in both fields is constant
upon such an exchange. In a case where $\Omega_p \sim \Omega_c$
one must take into account that it is the total field strength
$|\Omega|^2 = |\Omega_p|^2 + |\Omega_c|^2$, rather than simply the
input coupling strength $\Omega_{c0}^2$ which enters the numerator
in our equation for the group velocity $V_g(z)$ (\ref{eq:Vg}).
However, the expressions for the Rabi frequencies involve the
oscillator strengths of the relevant transitions. Thus, preserving
the total number of {\it photons} at each point in space will
result in a homogenous $|\Omega|^2 = \Omega_{c0}^2$ if and only if
$f_{13}=f_{23}$.  During the time that the probe is completely
contained in the BEC, $|\Omega|^2$ will differ from
$\Omega_{c0}^2$ in the vicinity of the pulse according to:

\begin{equation}
\label{eq:adiabaton} |\Omega(z,t)|^2 = \Omega_{c0}^2 +
|\Omega_p(z,t)|^2\bigg(1 - \frac{f_{23}}{f_{13}}\bigg)
\end{equation}

\begin{figure}
\includegraphics{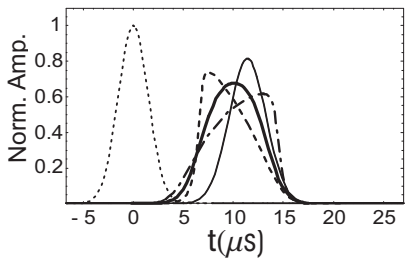} \caption{
\label{fig:strongProbe}\textbf{USL with a strong probe.}  The
dotted curve shows a strong probe pulse amplitude with peak value
$\Omega_{p0}=\Omega_{c0}=(2\pi)~8~\mathrm{MHz}$ input into a
sodium cloud with an optical density $D(z_\mathrm{out})=407$.  We
use an oscillator strength $f_{13}=1/2$ but vary the coupling
oscillator strength $f_{23}$. The output
$\Omega_p^\mathrm{(out)}/\Omega_{p0}$ is shown in the cases
$f_{23}=1/2$ (thick solid curve), $f_{23}=2/3$ (dot-dashed curve),
and $f_{23}=1/3$ (dashed curve). For reference, the thin solid
curve shows output for a weak input probe
$\Omega_{p0}=(2\pi)~1.4~\mathrm{MHz}$.}
\end{figure}

\noindent This leads to a nonlinear distortion, whereby the center
of the pulse (where the probe intensity is greatest) tends to move
faster or slower than the front and back edges, depending on the
sign of term in parentheses. This behavior is quite clearly seen
in Fig.~\ref{fig:strongProbe}, where we compare a delayed weak
probe regular USL output pulse to cases with strong probes
$\Omega_{p0}=\Omega_{c0}$ and vary $f_{23}=1/3,1/2,2/3$
($f_{13}=1/2$ in all cases).   Each time we chose the coupling
intensity so that that the Rabi frequency is $\Omega_{c0}=(2
\pi)~8~\mathrm{MHz}$. One sees that when $f_{13}=f_{23}$ there is
no asymmetrical distortion, though we get a slight spreading and a
shorter delay time. However, in the other two cases, there is
asymmetry in the output pulses, consistent with the picture that
the center of the pulse will travel slower or faster than the
edges according to (\ref{eq:adiabaton}) and (\ref{eq:Vg}) (with
$\Omega_{c0}^2$ replaced by $|\Omega|^2$ in the numerator).  The
result is a breakdown of our estimate (\ref{eq:probeOutIdeal}). As
this is a propagation effect, the absolute magnitude of this
effect should roughly be linear with optical density and the
relative distortion will go as the optical density divided by the
pulse length.  This is the primary reason it would be advantageous
to use the levels we chose in our simulation in
Fig.~\ref{fig:solitonOutput} if there is a strong probe.

We just outlined the reason that the equal oscillator strengths
can be important in reducing distortion during propagation and
thus $E_{\mathrm{out}}$.  It turns out that the writing process is
also more robust when $f_{13}=f_{23}$. The reason is the presence
of the additional term $\alpha_{12}$ (\ref{eq:alphaDef}). This
term is quite small in the weak probe limit, but in the strong
probe case leads to an additional phase shift, not accounted for
in (\ref{eq:probeRevivePhase}), which depends in detail on the
pulse amplitude and structure.

\begin{figure}
\includegraphics{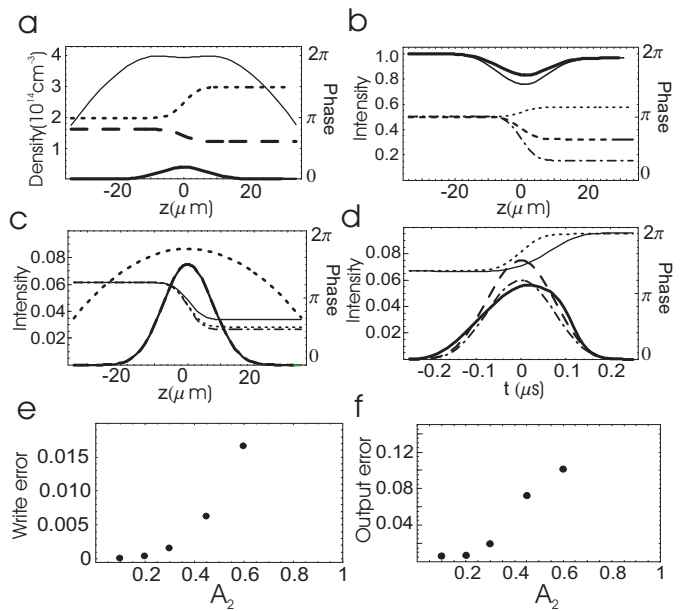}
\caption{\label{fig:test2}\textbf{Writing information with unequal
oscillator strengths.}  \textbf{(a-d)} Plots are for the same
parameters as Fig.~\ref{fig:test1} except with $f_{12}=1/12, \,
f_{23}=1/4$ and $A_2=0.25$. \textbf{(e-f)} Write $E_w$ and output
$E_{\mathrm{out}}$ errors versus $A_2$ for the parameters
$L_2=10~\mu\mathrm{m}, \, L_{\phi_1}=L_{\phi_2}= 5~\mu\mathrm{m},
\, A_{\phi_1}= 0.5 \, \pi, \,A_{\phi_2}= -0.2 \, \pi$.}
\end{figure}

Figs.~\ref{fig:test2}(a-d) shows a case, similar to
Fig.~\ref{fig:test1}, but with $f_{13}=1/12,f_{23}=1/4$ (as in the
cases in Figs.~\ref{fig:fringeEx}-\ref{fig:breatheEx2}), and with
a fairly small nonlinearity $A_2=0.25$. One sees in
Fig.~\ref{fig:test2}(b) that there is a dip in $\Omega_D$ in the
region of the probe due to the $\alpha_{12}$ term.
Fig.~\ref{fig:test2}(c) shows the written probe pulse.  There is a
small but discernable difference between the predicted
$\phi_p^\mathrm{(rev-ideal)}$ (\ref{eq:probeRevivePhase}) and
actual $\phi_p^\mathrm{(rev)}$ phase. Fig.~\ref{fig:test2}(d) then
shows the asymmetric distortion which develops as the strong probe
propagates out. One sees the phase jump is distorted in addition
to the amplitude. Figs.~\ref{fig:test2}(e-f) demonstrate how these
effects lead to substantially higher errors $E_w$ and
$E_{\mathrm{out}}$ as $A_2$ becomes larger.  From our estimate
from the previous section we expect an output error of
$E_{\mathrm{out}}=0.0147$ and we see in Fig.~\ref{fig:test2}(f)
that the distortion effect leads to errors higher than this when
$A_2 \ge 0.3$.

We thus conclude that fidelity of both the writing and output is
compromised in a system with unequal oscillator strengths.
However, this only comes into effect in cases where the
nonlinearity is important.  In weak probe cases neither of these
effects is important and the fidelity can still be estimated with
the analysis of the $f_{13}=f_{23}$ case in
Sections~\ref{subsec:fidelity}-\ref{subsec:error}.

\section{\label{sec:conclusion} Outlook}

In conclusion, we have established first several new results
regarding the fast switching of the coupling field in light
storage experiments. We found the switching could be done
arbitrarily fast without inducing absorptions so long as the probe
is completely contained in the atomic medium (and $V_g \ll c$
which is usually the case in practice).  We have also seen that
when the switching is slow compared to the excited state lifetime
($\sim 25$~ns), the probe smoothly follows the temporal switching
of the coupling field, while in the other limit the probe
amplitude undergoes oscillations which damp out with this time
scale (see Fig.~\ref{fig:fastSwitching}).

Next, we saw that these stopped pulses can induce novel and rich
two-component BECs dynamics during the storage time.  Both the
relative scattering lengths of the states used and the probe to
coupling intensity ratio have a strong effect on the qualitative
features of the dynamics.  In the weak probe case, the $|2
\rangle$ condensate sees en either an effective repulsive hill
(when $a_{12}>a_{11}$, see Fig.\ref{fig:fringeEx}(b)) or harmonic
oscillator potential (when $a_{12}<a_{11}$, see
Fig.\ref{fig:breatheEx2}(a)). The characteristic timescale for the
dynamics is given my the chemical potential of the BEC (in our
cases $\sim (2 \pi)~1~$kHz) times the relative scattering length
difference $|a_{12}-a_{11}|/a_{11}$.   In the latter case, the
resulting evolution can be easily calculated by decomposing the
pulse into the various harmonic oscillator eigenstates.  Thus it
is possible to choose input pulses which preserve their density
over time or undergo predictable reshaping such as dipole sloshing
or breathing, allowing controlled storage and linear processing of
optical information. Inputting stronger probes will add
nonlinearity to the evolution, making nonlinear processing
possible.  Very strong probes and phase separating scattering
lengths $a_{12}^2 > a_{11} a_{22}$ lead to the formation and
motion of vector solitons, which have not been observed in BECs to
date.

We then showed that switching the coupling field on after the
dynamics writes the various density and phase features of the
wavefunctions onto revived probe pulses.  This was seen
qualitatively for various examples
(Figs.~\ref{fig:fringeEx}-\ref{fig:solitonOutput}).  A precise
relationship between the wavefunctions and output pulses was found
(\ref{eq:probeReviveMag})-(\ref{eq:probeOutIdeal}), meaning the
output probe pulse can be used as a diagnostic of the relative
density and phase in the BEC wavefunctions. We have also
identified sources of attenuation and distortion in the writing
and output processes and quantitative errors were calculated for a
wide range of length and amplitudes of Gaussian pulses (Figs.
\ref{fig:seriesLength}-\ref{fig:seriesAmp}). As we saw there, the
error during the output dominates for shorter pulses and is
$(5/16)\beta^2$ (see Eq.~(\ref{eq:Tspace})).  For longer pulses
effects due to the condensate edge introduce errors into the
writing process. Balancing these two considerations one can
optimize the fidelity, which improves with optical density.

For the strong probe case, we found that for equal oscillator
strengths ($f_{13} = f_{23}$), one could still relate the
wavefunctions to the output pulses by taking into account an
additional nonlinear phase shift (\ref{eq:probeRevivePhase}). The
fidelity of transfer of information was virtually independent of
the probe strength $\Omega_{p0}^2$ even when $\Omega_{p0} \sim
\Omega_{c0}$ (see Fig.~\ref{fig:seriesAmp}).  For unequal
oscillator strengths ($f_{13} \not= f_{23}$) we found strong
probes lead to additional features in the phase pattern during the
writing process and in distortion during the propagation during
the output, leading to much higher errors (see
Fig.~\ref{fig:test2}).

Looking towards future work, we note that it is trivial to extend
the analysis here to cases where there are two or more spatially
distinct BECs present. So long as they are optically connected,
information contained in the form of excitations of one BEC could
be output with this method and then input to another nearby BEC,
leading to a network. In the near future, there is the exciting
possibility of extending these results beyond the mean field to
learn how using this method of writing onto light pulses could be
used as a diagnostic of quantum evolution in BECs. A particular
example of interest would be to investigate the spin squeezing due
to atom-atom interactions in a two-component system and the
subsequent writing of the squeezed statistics onto the output
probe pulses.  Furthermore, we expect that performing revival
experiments after long times could be used as a sensitive probe of
decoherence in BEC dynamics, similar to the proposal in
\cite{decoherence}. Lastly, there is the prospect of inputting two
or more pulses in a BEC and using controllable nonlinear
processing from atom-atom interactions to design of multiple bit
gates, such as conditional phase gates.  We anticipate the results
presented here can be applied to these problems as well as to
applications in quantum information storage, which require the
ability to transfer coherent information between light and atom
fields.

\appendix*

\section{\label{app:adElim} Adiabatic elimination of $\psi_3$ and faster switching}

We arrived at Eqs.~(\ref{eq:formalism12})-(\ref{eq:formalismPC})
by first considering a system of equations with all three levels
considered and then adiabatically eliminating the wavefunction for
$| 3 \rangle$.  For completeness we write the original equations
here.

The three coupled GP equations are:

\begin{eqnarray}
\label{eq:fullGP1}  i \hbar \frac{\partial \psi_1 }{\partial t} &
= & \bigg[-\frac{\hbar^2}{2m}\frac{\partial^2}{\partial z^2} +
V_1(z) + U_{11} | \psi_1 |^2   +  U_{12}| \psi_2 |^2 \bigg] \psi_1 \nonumber \\
& & \hspace{1cm} + \frac{1}{2}\hbar \, \Omega_p^* \psi_3, \\
\label{eq:fullGP2} i \hbar \frac{\partial \psi_2}{\partial t} & =
& \bigg[-\frac{\hbar^2}{2m}\frac{\partial^2}{\partial z^2} +
V_2(z) + U_{22}
|\psi_2|^2 +  U_{12}| \psi_1 |^2 \bigg] \psi_2 \nonumber \\
& & \hspace{1cm} + \frac{1}{2}\hbar \, \Omega_c^* \psi_3, \\
\label{eq:fullGP3} i \hbar \frac{ \partial \psi_3}{\partial t}  &
= &
 \frac{1}{2} \Omega_p \hbar \, \psi_1  + \frac{1}{2}  \Omega_c \hbar \psi_2 - i \frac{\Gamma}{2} \psi_3
\end{eqnarray}

\noindent Note we have ignored the external dynamics of $\psi_3$
altogether as they will be negligible compared with $\Gamma,
\Omega_p, \Omega_c$ (on the order of three orders of magnitude
slower). Maxwell's equations read:

\begin{eqnarray}
\label{eq:Maxwell}  \left( \frac{\partial}{\partial z}+
\frac{1}{c}\frac{\partial}{\partial t} \right) \Omega_p & = & -
i \frac{f_{13} \sigma_0}{A} \frac{\Gamma}{2} N_c \psi_3 \psi_1^*, \nonumber \\
 \left( \frac{\partial}{\partial z}+
\frac{1}{c}\frac{\partial}{\partial t} \right) \Omega_c & = & - i
\frac{f_{23} \sigma_0}{A} \frac{\Gamma}{2} N_c \psi_3 \psi_2^*.
\end{eqnarray}

\noindent In adiabatically eliminating $\psi_3$, we assume all
quantities vary slowly compared to excited state lifetime
$\Gamma^{-1}$ ($\sim 16$~ns in sodium).  This allows us to set $d
\psi_3/dt \rightarrow 0$ in (\ref{eq:fullGP3}) and arrive at:

\begin{eqnarray}
\label{eq:psi3} \psi_3 \approx
 - \frac{i}{\Gamma} \big(\Omega_p \psi_1 + \Omega_c \psi_2\big)
\end{eqnarray}

\noindent Plugging this into (\ref{eq:fullGP1})-(\ref{eq:fullGP2})
and (\ref{eq:Maxwell}) yields
(\ref{eq:formalism12})-(\ref{eq:formalismPC}).

\begin{acknowledgments}
This work was supported by the U.S. Air Force Office of Scientific
Research, the U.S. Army Research Office OSD Multidisciplinary
University Research Initiative Program, the National Science
Foundation, and the National Aeronautics and Space Administration.
The authors would like to thank Janne Ruostekoski for helpful
discussions.
\end{acknowledgments}


\begin{thebibliography}{99}

\bibitem{quantumInfo} David. P. DiVincenzo, Fortsch. Phys. {\bf 48} 771 (2000).

\bibitem{Nature1} L.V. Hau, S.E. Harris, Z. Dutton, and C.H. Behroozi, Nature
{\bf 397}, 594 (1999).

\bibitem{OtherUSL} M.M. Kash, {\it et al.}, Phys. Rev. Lett.
 {\bf 82} 5229 (1999); D. Budker, D.F. Kimball, S.M. Rochester, and
V.V. Yaschuk, Phys. Rev. Lett. {\bf 83}, 1767 (1999).

\bibitem{Nature2} C. Liu, Z. Dutton, C.H. Behroozi, and L.V. Hau, Nature {\bf
409}, 490 (2001).

\bibitem{OtherStoppedLight} D.F. Phillips, A. Fleischhauer, A. Mair, R.L.
Walsworth, and M.D. Lukin, Phys. Rev. Lett. {\bf 86}, 783 (2001);
A. Mair, J. Hager, D.F. Phillips, R.L. Walsworth, and M.D. Lukin,
Phys. Rev. A {\bf 65}, 031802 (2002).

\bibitem{quantumProcessing} M.D. Lukin, S.F. Yelin, and M.
Fleischhauer, Phys. Rev. Lett.  {\bf 84}, 4232 (2000).

\bibitem{EIT} S.E. Harris, Physics Today {\bf 50}, 36
(1997).

\bibitem{HauTalks} L.V. Hau, ``BEC and Light Speeds of 38
miles/hour'', Feb. 10, 1999, Workshop on Bose-Einstein
Condensation and Degenerate Fermi Gases, Center for Theoretical
Atomic, Molecular, and Optical Physics, (Boulder, CO),
http://fermion.colorado.edu/$\sim$chg/Talks/Hau; L.V. Hau,
``Bose-Einstein Condensation and light speeds of 38 miles per
hour'', Harvard University Physics Dept. Colloquium, Feb. 22,
1999, videocasette (Harvard University, Cambridge, MA, 1999).

\bibitem{SolitonVortex} Z. Dutton, M. Budde, C. Slowe, and L.V.
Hau, Science {\bf 293}, 663 (2001); Science Express, published
online 28 June 2001, 10.1126/science.1062527.

\bibitem{BEC}  M. Inguscio, S. Stringari, and C.
Wieman, eds., {\it Bose-Einstein Condensates in Atomic Gases,
Proceedings of the International School of Physics Enrico Fermi,
Course CXL}, (International Organisations Services B.V.,
Amsterdam, 1999).

\bibitem{stoppedLight} M. Fleischhauer and M.D. Lukin,
Phys. Rev. Lett. {\bf 84}, 5094 (2000).

\bibitem{fastStopped} A.B. Matsko, {\it et al.}, Phys. Rev. A,
{\bf 64} 043809 (2001).

\bibitem{stoppedTheory} M. Fleischhauer and M.D. Lukin,
Phys. Rev. A {\bf 65}, 022314 (2002).

\bibitem{strongStopped} T.N. Dey and G.S. Agarwal, Phys. Rev. A
{\bf 67}, 033813 (2003).

\bibitem{inelastic} C.J. Myatt,  E.A. Burt, R.W. Ghrist, E.A. Cornell,
and C.E. Weiman, Phys. Rev. Lett. {\bf 78}, 586 (1997); J.P.
Burke, C.H. Greene, J.L Bohn, Phys. Rev. Lett. {\bf 81}, 3355
(1998). A Mathematica notebook is available at
http://fermion.colorado.edu/$\sim$chg/Collisions/.

\bibitem{multipleComponentJILA} D. S. Hall, M. R. Matthews, C. E. Wieman, and E. A.
Cornell, Phys. Rev. Lett. {\bf 81}, 1539 (1998); {\it ibid.} {\bf
81}, 1543 (1998).

\bibitem{twoCompSolitons} S.V. Manakov, Soviet JETP {\bf 465} 505 (1973);
Th. Busch and J.R. Anglin, Phys. Rev. Lett. {\bf 87}, 101401
(2001).

\bibitem{HarrisHau} S.E. Harris and L.V. Hau, Phys. Rev. Lett.
{\bf 82}, 4611 (1999).

\bibitem{FiniteTemp} S. Giorgini, L.P. Pitaevskii, and S. Stringari, J.
Low Temp. Phys. {\bf 109}, 309 (1997).

\bibitem{adiabatic} J. Javanainen and J. Ruostekoski, Phys. Rev.
A {\bf 52} 3033 (1995).

\bibitem{thesis} Z. Dutton, Ph.D. thesis (Harvard University,
2002).

\bibitem{numerics} The code uses a Crank-Nicolson technique for
the external propagation of the atomic fields and a symmmetric
second order propagation for the light field internal coupling
terms.  The light fields are propagated in space with a
second-order Runge-Kutta algorithm.  The time steps varied between
$dt=4$~ns during the light propogation to as high as
$dt=2.5~\mu\mathrm{s}$ during the storage time with the light
fields off.  A spatial grid with typical spacing
$dz=0.1~\mu\mathrm{m}$ was used.  The intervals $dt,dz$ were
varied to assure they did not effect the results.

\bibitem{imaginaryTime} J.E. Williams, Ph.D. thesis (University of
Colorado, 1999).

\bibitem{TF} G. Baym and C. Pethick, Phys. Rev. Lett. {\bf 76}, 6
(1996).

\bibitem{scatteringLengthNa}  E. Tiesinga, {\it et al.},
J. Res. Natl. Inst. Stand. Techol. {\bf 101}, 505 (1996).

\bibitem{darkState} E. Arimondo and G. Orriols, Nuovo Cimento {\bf
17}, 333 (1976).

\bibitem{adiabatons} R. Grobe, F.T. Hioe, and J.H. Eberly, Phys. Rev. Lett.
{\bf 73}, 3183 (1994).

\bibitem{normalModes} S.E. Harris, Phys. Rev. Lett. {\bf 72}, 52
(1994).

\bibitem{numerics3Comp} In this code, described in \cite{thesis}, we keep track of the amplitudes
$\psi_1,\psi_2,\psi_3$ and propagate them and $\Omega_p,\Omega_c$
according to the analogues of
Eqs.~(\ref{eq:fullGP1})-(\ref{eq:Maxwell}).  We do not include the
atomic dynamics in this code, which are on a much slower time
scale.

\bibitem{scatteringLengthRb} J.M. Vogels, R.S. Freeland, C.C. Tsai,
B.J. Verhaar, and D.J. Heinzen, Phys. Rev. A {\bf 61} 043407
(2002).

\bibitem{phaseSeperation} H.-J. Miesner {\it et al.}, Phys. Rev. Lett.
{\bf 82}, 2228 (1999).

\bibitem{phaseSeperationTh} T-L. Ho and V.B. Shenoy, Phys. Rev. Lett.
{\bf 77}, 3297 (1996); H. Pu and N.P. Bigelow, Phys. Rev. Lett.
{\bf 80}, 1130 and 1134 (1998); P. Ao and S.T. Chui, Phys. Rev. A
{\bf 58}, 4836 (1999).

\bibitem{feshbach} S. Inouye {\it et al.}, Nature {\bf 392},
151 (1998);  S.L. Cornish {\it et al.}, Phys. Rev. Lett. {\bf 85},
1795-1798 (2000).

\bibitem{opticalTrap} D.~M. Stamper-Kurn, {\it et
al.}, Phys. Rev. Lett. {\bf 80}, 2027-2030 (1998).

\bibitem{breatheTogether} A. Sinatra and Y. Castin, European Phys.
Jour. D {\bf 8}, 319 (2000).

\bibitem{snake} B.B. Kadomtsev and V.I. Petviashvili, Sov.
Phys. Dokl. {\bf 15}, 539 (1970); D.L. Feder, {\it et al.}, Phys.
Rev. A {\bf 62}, 053606 (2000); B.P. Anderson, {\it et al.}, Phys.
Rev. Lett. {\bf 86}, 2926 (2001).

\bibitem{spinSqueezing} M. Kitagawa and M. Ueda,
Phys. Rev. A {\bf 47}, 5138 (1993); A. Sorensen, L.-M. Duan, J.I.
Cirac, and P. Zoller, Nature {\bf 409}, 63 (2001);  U.V. Poulson
and K. Molmer, Phys. Rev. A, {\bf 64}, 013616 (2001).

\bibitem{squeezedLight} {\it Quantum Noise Reduction in Optical
Systems}, eds. C. Fabre and E. Giacobino [Appl. Phys. B {\bf 55},
279 (1992)].

\bibitem{squeezedLightRaman}  U.V. Poulson and K. Molmer, Phys. Rev. Lett.  {\bf 87},
123601 (2001).

\bibitem{decoherence} J. Ruostekoski and D.F. Walls, Phys. Rev. A
{\bf 59} R2571 (1999).

\end{thebibliography}
\end{document}